\documentclass[number,seceqn,secthm,citesort,dvips]{arxstspdf}
\usepackage{mathrsfs}
\usepackage{graphicx}
\usepackage{flushend}
\usepackage{stfloats}

\volume{23}
\issue{3}
\pubyear{2008}
\firstpage{365}
\lastpage{382}
\doi{10.1214/08-STS265}

\makeatletter
\newproclaim{ex}[thm]{Example}
\newproclaim{defn}[thm]{Definition}
\newproclaim{remark}[thm]{Remark}
\newtheorem{cor}[thm]{Corollary}
\newtheorem{prop}[thm]{Proposition}

\newtheorem{lem}[thm]{Lemma}

\newproclaim{samcalg}{SAMC Algorithm}
\newproclaim{amalgg}{AM Algorithm}
\newproclaim{saemalg}{SAEM Algorithm}
\newproclaim{newtonalg}{Newton's Algorithm}
\newproclaim{NPalg}{N$+$P Algorithm}

\def\eqref#1{(\ref{#1})}
\makeatother

\begin{document}
\begin{frontmatter}

\title{Stochastic Approximation and Newton's Estimate of a Mixing Distribution}
\runtitle{SA and Newton's Algorithm}

\begin{aug}
\author[a]{\fnms{Ryan} \snm{Martin}\corref{}\ead[label=e1]{martinrg@stat.purdue.edu}} \and
\author[b]{\fnms{Jayanta K.} \snm{Ghosh}\ead[label=e2]{ghosh@stat.purdue.edu}}
\runauthor{R. Martin and J. K. Ghosh}

\affiliation{Purdue University and Indian Statistical Institute}

\address[a]{Ryan Martin is Graduate Student,
Department of Statistics, Purdue University, 250 North University Street,
West Lafayette, Indiana 47907, USA, \printead{e1}.}
\address[b]{Jayanta~K. Ghosh is Professor, Department of Statistics, Purdue University,
150 N. University Street, West Lafayette, 47907, USA and Professor
Emeritus, Division of Theoretical
Statistics and Mathematics,
Indian Statistical Institute, 203 B T Road Kolkata, India 700108,
\printead{e2}.}
\end{aug}

%
\begin{abstract}
Many statistical problems involve mixture models and the need for
computationally
efficient methods to estimate the mixing distribution has increased dramatically
in recent years. Newton [\textit{Sankhy\=a Ser. A} \textbf{64} (2002)
306--322]
proposed a fast recursive algorithm
for estimating the mixing distribution, which we study as a special
case of
stochastic approximation (SA). We begin with a review of SA, some recent
statistical applications, and the theory necessary for analysis of a SA
algorithm, which includes Lyapunov functions and ODE stability theory.
Then standard SA results are used to prove consistency of Newton's estimate
in the case of a finite mixture. We also propose a modification of Newton's
algorithm that allows for estimation of an additional unknown parameter in
the model, and prove its consistency.
\end{abstract}

%

\end{frontmatter}

\section{Introduction}
\label{S:intro}

The aim of the present paper is to review the subject of stochastic
approximation (SA), along the way highlighting some recent statistical
applications, and to explore its relationship with a recent
algorithm~\cite{nqz,newtonzhang,newton} for estimating a mixing distribution.

SA was introduced in~\cite{robbinsmonro} as an algorithmic method for
finding the root of a function $h$ when only noisy observations on $h$
are available. It has since developed into an important area of systems
control and optimization, with numerous applications in statistics. In
Section~\ref{S:sa} we give a brief introduction to the SA algorithm
and review three recent and innovative statistical applications. The
first two \cite{saem,haario} strengthen the EM and Metropolis
algorithms, respectively, and the third is a versatile Monte Carlo
integration method, called Stochastic Approximation Monte Carlo
(SAMC)~\cite{samc}, which can be applied in a variety of statistical
problems. We demonstrate that combining SAMC with the
\textit{energy--temperature duality}
\cite{kzw} provides a
method for estimating the normalizing constant of a density. We then
state a theorem providing sufficient conditions for almost sure
convergence of a SA algorithm, which is used in
Section~\ref{S:mixingestimate} to study the convergence properties of a mixing
distribution estimate. For this purpose, the necessary stability theory
for ordinary differential equations (ODEs) is developed.

Many statistical problems involve modeling with latent, or unobserved,
random variables, for example, cluster analysis~\cite{mclachlan} and multiple
testing or estimation with high-dimensional data
\cite{allison,efrontibs,efron2001,scottberger,tang}. The distribution of the
manifest, or observed, random variables then becomes a mixture of the form
%
\begin{equation}\label{E:marginal}
\Pi_f(x) = \int_{\Theta} p(x|\theta) f(\theta) \,d\mu(\theta),
\end{equation}
where $\theta\in\Theta$ is the latent variable or parameter, and $f$
is an unknown mixing density with respect to the measure $\mu$ on
$\Theta$. Estimation of $f$ plays a fundamental role in many inference
problems, such as an empirical Bayes approach to multiple testing.

For the deconvolution problem, when $p(x|\theta)$ in \eqref
{E:marginal} is of the form $p(x-\theta)$, asymptotic results for
estimates of $f$, including optimal rates of convergence, are
known~\cite{fan}. A nonparametric Bayes approach to Gaussian
deconvolution is discussed in~\cite{ghosh-ram}. For estimating $\Pi
_f$, a Bayesian might assume an a priori distribution on $f$,
inducing a prior on $\Pi_f$ via the map $f \mapsto\Pi_f$.
Consistency of the resulting estimate of $\Pi_f$ is considered
in~\cite{ggr,barron,ghosal-van}.

In Section~\ref{S:mixingestimate}, we describe a recursive algorithm
of Newton et al.~\cite{nqz,newtonzhang,newton} for estimating the
mixing density $f$. This estimate is significantly faster to compute
than the popular nonparametric Bayes estimate based on a Dirichlet
process prior. In fact, the original motivation~\cite{nqz} for the
algorithm was to approximate the computationally expensive Bayes
estimate. The relative efficiency of the recursive algorithm compared
to MCMC methods used to compute the Bayes estimate, coupled with the
similarity of the resulting estimates, led Quintana and Newton~\cite
{quintana2000} to suggest the former be used for \emph{Bayesian
exploratory data analysis}.

While Newton's algorithm performs well in examples and simulations
(see \cite{nqz,newtonzhang,quintana2000,newton,ghosh,tmg} and
Section~\ref{SS:examples}), very little is known about its
large-sample properties. A rather difficult proof of consistency, based
on an approximate martingale representation of the Kullback--Leibler
divergence, is given by Ghosh and Tokdar~\cite{ghosh} when $\Theta$
is finite; see Section~\ref{SS:newton-review}. In
Section~\ref{SS:newtonconvsa}, we show that Newton's algorithm can be expressed as
a stochastic approximation and results presented in
Section~\ref{SS:sa-thm} are used to prove a stronger
consistency theorem than
in~\cite{ghosh} for the case of finite~$\Theta$, where the
Kullback--Leibler divergence serves as the Lyapunov function.

The numerical investigations in Section~\ref{SS:examples} consider two
important cases when $\Theta$ is finite, namely, when $f$~is strictly
positive on $\Theta$ and when $f(\theta)=0$ for some $\theta\in
\Theta$. In the former case, our calculations show that Newton's
estimate is superior, in terms of accuracy and computational
efficiency, to both the nonparametric MLE and the Bayes estimate. For
the latter case, when only a superset of the support of $f$ is known,
the story is completely different. While Newton's estimate remains
considerably faster than the others, it is not nearly as accurate.

We also consider the problem where the sampling density $p(x|\theta)$
of \eqref{E:marginal} is of the form $p(x|\theta,\xi)$, where $f$ is
a mixing density or prior for $\theta$, and $\xi$ is an additional
unknown parameter. Newton's algorithm is unable to handle unknown
$\xi$, and we propose a modified algorithm, called N${}+{}$P, capable of
recursively estimating both $f$ and $\xi$. We express this algorithm
as a general SA and prove consistency under suitable conditions.

In Section \ref{S:discuss} we briefly discuss some additional
theoretical and practical issues concerning Newton's recursive
algorithm and the N${}+{}$P.

\section{Stochastic Approximation}\label{S:sa}

\subsection{Algorithm and Examples}\label{SS:sa-review}

Consider the problem of finding the unique root $\xi$ of a function
$h(x)$. If $h(x)$ can be evaluated exactly for each $x$ and if $h$ is
sufficiently smooth, then various numerical methods can be employed to
locate $\xi$. A~majority of these numerical procedures, including the
popular Newton--Raphson method, are iterative by nature, starting with
an initial guess $x_0$ of $\xi$ and iteratively defining a sequence
$\{x_n\}$ that converges to $\xi$ as $n \to\infty$. Now consider the
situation where only noisy observations on $h(x)$ are available; that
is, for any input $x$ one observes $y=h(x)+\varepsilon$, where
$\varepsilon$ is a
zero-mean random error. This problem arises in situations where $h(x)$
denotes the expected value of the response when the experiment is run
at setting $x$. Unfortunately, standard deterministic methods cannot be
used in this problem.

In their seminal paper, Robbins and Monro~\cite{robbinsmonro} proposed
a \emph{stochastic approximation} algorithm for defining a sequence of
design points $\{x_n\}$ targeting the root $\xi$ of $h$ in this noisy
case. Start with an initial guess $x_0$. At stage $n \geq1$, use the
state $x_{n-1}$ as the input, observe $y_n = h(x_{n-1}) + \varepsilon_n$, and
update the guess $(x_{n-1},y_n) \mapsto x_n$. More precisely, the
Robbins--Monro algorithm defines the sequence $\{x_n\}$ as follows:
start with $x_0$ and, for $n \geq1$, set
%
\begin{eqnarray}\label{E:r-m}
x_n & =& x_{n-1} + w_n y_n
\nonumber\\[-8pt]
\\[-8pt]
& =& x_{n-1} + w_n \{h(x_{n-1}) + \varepsilon_n \},
\nonumber
\end{eqnarray}
where $\{\varepsilon_n\}$ is a sequence of i.i.d. random variables with mean
zero, and the weight sequence $\{w_n\}$ satisfies
%
\begin{equation}\label{E:sequence}
w_n > 0, \quad\sum_n w_n = \infty, \quad\sum_n w_n^2 < \infty.
\end{equation}

While the SA algorithm above works in more general situations, we can
develop our intuition by looking at the special case considered
in~\cite{robbinsmonro}, namely, when $h$ is bounded, continuous and
monotone decreasing. If $x_n < \xi$, then $h(x_n) > 0$ and we have
\begin{eqnarray*}
\mathbb{E}(x_{n+1}|x_n) & =& x_n + w_{n+1}
\{h(x_n) + \mathbb{E}(\varepsilon_{n+1})\}
\\
& =& x_n + w_{n+1}h(x_n)
\\
& >& x_n.
\end{eqnarray*}
Likewise, if $x_n > \xi$, then $\mathbb{E}(x_{n+1}|x_n) < x_n$. This shows
that the move $x_n \mapsto x_{n+1}$ will be in the correct direction
\emph{on average}.

Some remarks on the conditions in \eqref{E:sequence} are in order.
While $\sum_n w_n^2 < \infty$ is necessary to prove convergence, an
immediate consequence of this condition is that $w_n \to0$. Clearly
$w_n \to0$ implies that the effect of the noise vanishes as $n \to
\infty$. This, in turn, has an averaging effect on the iterates $y_n$.
On the other hand, the condition $\sum_n w_n = \infty$ washes out the
effect of the initial guess $x_0$. For further details, see \cite{nevelson}.

\begin{figure}

\includegraphics{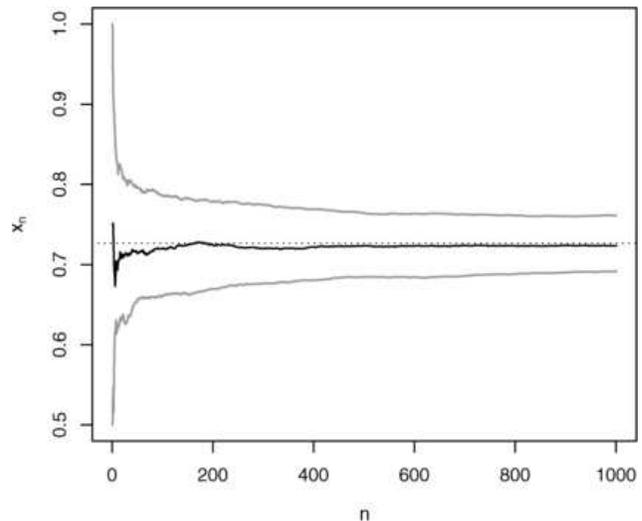}

\caption{Sample paths of the three SA sequences $\{x_n\}$ in
Example~\protect\ref{EX:t-quantile}. The dotted line is the exact 75th
percentile of the $t_5$ distribution.\label{F:t-quantile}}
\end{figure}

\begin{figure}

\includegraphics{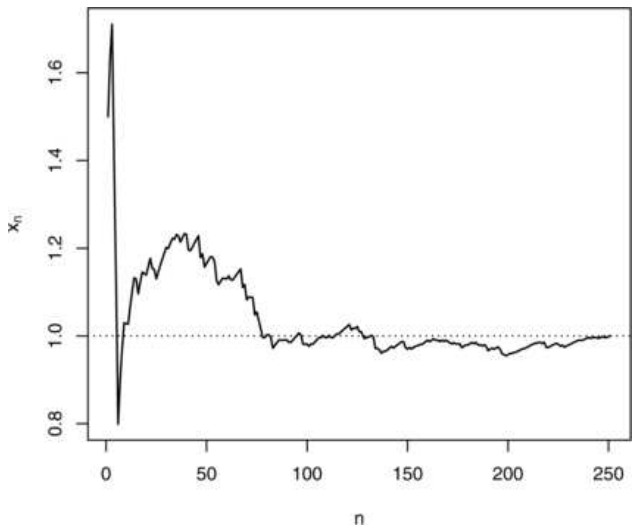}

\caption{Sample path of the sequence $\{x_n\}$ in Example~\protect\ref{EX:eb}.
The dotted line is the value of $\xi$ used for data
generation.\label{F:eb-sa}}
\end{figure}

We conclude this section with three simple examples of SA to shed light
on when and how the algorithm works. Example~\ref{EX:first-ex}, taken
from \cite{kushner}, page~4, is an important special case of the
Robbins--Monro algorithm~\eqref{E:r-m} which further motivates the
algorithm as well as the conditions~\eqref{E:sequence} on the sequence
$\{w_n\}$. Example \ref{EX:t-quantile} uses SA to find quantiles of a
$t$-distribution, and Example \ref{EX:eb} illustrates a connection
between SA and \textit{empirical Bayes}, two of Robbins's greatest contributions.
\begin{ex}\label{EX:first-ex}
Let $F_\xi$ be the cdf of a distribution with mean $\xi$. Then
estimation of $\xi$ is equivalent to solving $h(x)=0$ where
$h(x) =\xi-x$. If
$Z_1,\ldots,Z_n$ are i.i.d. observations from $F_\xi$, then the
average $\overline{Z}_n$
is the least squares estimate of~$\xi$. To see that $\{\overline{Z}_n\}$ is
actually a SA sequence, recall the computationally efficient recursive
expression for~$\overline{Z}_n$:
%
\begin{equation}\label{E:zbar}
\overline{Z}_n = \overline{Z}_{n-1} + n^{-1}(Z_n - \overline{Z}_{n-1}).
\end{equation}
If we let $x_n = \overline{Z}_n$, $w_n = n^{-1}$
and $y_n = Z_n-\overline{Z}_{n-1}$, then \eqref{E:zbar}
is exactly of the form of \eqref{E:r-m},
with $\{w_n\}$ satisfying \eqref{E:sequence}. Moreover, if
$\varepsilon_n =
Z_n-\xi$, then we can write $y_n = h(x_{n-1})+\varepsilon_n$. With this
setup, we could study the asymptotic behavior of $x_n$ using the SA
analysis below (see Sections~\ref{SS:ode} and \ref{SS:sa-thm}), although
the SLLN already guarantees $x_n \to\xi$ a.s.
\end{ex}
\begin{ex}\label{EX:t-quantile}
Suppose we wish to find the $\alpha$th quantile of the
$t_\nu$ distribution; that is, we want to find the solution to the
equation $F_\nu(x) = \alpha$, where $F_\nu$ is the cdf of the $t_\nu
$ distribution. While there are numerous numerical methods available
(e.g., Newton--Raphson or bijection), we demonstrate below how SA can be
used to solve this problem. Making use of the well-known fact that the
$t_\nu$ distribution is a scale-mixture of normals, we can write
\[
F_\nu(x) = \mathbb{E} [ \Phi( x | \nu^{-1}Z )  ],
\quad Z \sim\chi_\nu^2,
\]
where $\Phi(x|\sigma^2)$ is the cdf of the $N(0,\sigma^2)$
distribution. Now, for
$Z_1,Z_2, \ldots\stackrel{\mathrm{i.i.d.}}{\sim}\chi_\nu^2$, the
sequence $\{y_n\}$ defined as $y_n = \alpha- \Phi( x_{n-1}|\nu^{-1}Z_n)$
are noisy observations of $h(x_{n-1})=\alpha-F_\nu(x_{n-1})$.
This $h$ is bounded, continuous and monotone decreasing so
the Robbins--Monro theory says that the sequence $\{x_n\}$ defined as
\eqref{E:r-m} converges to the true quantile, for any initial
condition $x_0$. For illustration, Figure \ref{F:t-quantile} shows the
first 1000 iterations of the sequence $\{x_n\}$ for $\alpha= 0.75$,
$\nu= 5$ and for three starting values $x_0 \in\{0.5,0.75,1.0\}$.
\end{ex}
%
\begin{ex}\label{EX:eb}
In Section~\ref{S:mixingestimate} we consider a particular recursive
estimate and show that it is of the form of a general SA. It turns out
that the problem there can also be expressed as an \textit{empirical
Bayes} (EB) problem~\cite{robbins}. In this simple example, we
demonstrate the connection between SA and EB, both of which are
theories pioneered by Robbins. Consider the simple hierarchical model
\[
\lambda_1,\ldots,\lambda_n \stackrel{\mathrm{i.i.d.}}{\sim}
\operatorname{Exp}(\xi) \quad\mbox{and}\quad
Z_i|\lambda_i \stackrel{\mathrm{ind}}{\sim}
\operatorname{Poi}(\lambda_i)
\]
for $i=1,\ldots,n$, where the exponential rate $\xi> 0$ is unknown.
EB tries to estimate $\xi$ based on the observed data
$Z_1,\ldots,Z_n$. Here we consider a recursive estimate of $\xi$. Fix an initial
guess $x_0$ of $\xi$. Assuming $\xi$ is equal to $x_0$, the posterior
mean of $\lambda_1$ is $(Z_1+1)/(x_0+1)$, which is a good estimate of
$\xi^{-1}$ if $x_0$ is close to $\xi$. Iterating this procedure, we
can generate a sequence
%
\begin{equation}\label{E:eb-sa}
x_i = x_{i-1} + w_i  \biggl[ \frac{1}{x_{i-1}}
- \frac{Z_i+1}{x_{i-1}+1}  \biggr],
\end{equation}
where $\{w_i\}$ is assumed to satisfy \eqref{E:sequence}. Let $y_i$
denote the quantity in brackets in \eqref{E:eb-sa} and take its
expectation with respect to the distribution of $Z_i$:
%
\begin{equation}\label{E:eb-sa2}
h(x) = \mathbb{E}(y_i|x_{i-1}=x) = \frac{\xi-x}{\xi x (x+1)}.
\end{equation}
Then the sequence $\{x_n\}$ in \eqref{E:eb-sa} is a SA targeting a
solution of $h(x)=0$. Since $h$ is continuous, decreasing and $h(x)=0$
iff $x=\xi$, it follows from the general theory that $x_n \to\xi$.
Figure \ref{F:eb-sa} shows the first 250 steps of such a sequence with
$x_0 = 1.5$.
\end{ex}

The examples above emphasize one important property that $h(x)$ must
satisfy, namely, that it must be easy to ``sample'' in the sense that
there is a function $H(x,z)$ and a random variable $Z$ such that $h(x)
= \mathbb{E}[H(x,Z)]$. Another thing, which is not obvious from the examples,
is that $h(x)$ must have certain \emph{stability} properties. In
general, a SA sequence need not have a unique limit point. However,
conditions can be imposed which guarantee convergence to a particular
solution $\xi$ of $h(x)=0$, provided that $\xi$ is a \emph{stable}
solution to the ODE $\dot{x} = h(x)$. This is discussed further in
Section~\ref{SS:ode}.

\subsection{Applications}\label{SS:applications}

\subsubsection{\texorpdfstring{Stochastic approximation EM.}{Stochastic approximation EM}}\label{SSS:SAEM}

The EM algorithm~\cite{DLR} has quickly become one of the most popular
computational techniques for likelihood estimation in a host of
standard and nonstandard statistical problems. Common to all problems
in which the EM can be applied is a notion of ``missing data.''

Consider a problem where data $Y$ is observed and the goal is to
estimate the parameter $\theta$ based on its likelihood function
$L(\theta)$. Suppose that the observed data $Y$ is \emph{incomplete}
in the sense that there is a component $Z$ which is
\emph{missing}---this could be actual values which are not observed, as in
the case of censored data, or it could be latent variables, as in a
random effects model. Let $X=(Y,Z)$ denote the \emph{complete} data.
Then the likelihood function $f(z,\theta)$ based on the complete data
$x$ is related to $L(\theta)$ according to the formula $L(\theta) =
\int f(z,\theta)\,dz$. The EM algorithm produces a convergent sequence
of estimates by iteratively filling in the missing data $Z$ in the
E-step and then maximizing the simpler complete-data likelihood
function $f(z,\theta)$ in the M-step. The E-step is performed by
sampling the $z$-values from the density
\[
p(z|\theta) =\cases{%
f(z,\theta)/L(\theta), & if $L(\theta) \neq0$, \cr
0, & if $L(\theta) = 0$,}
\]
which is the predictive density of $Z$, given $Y$ and $\theta$.

It is often the case that at least one of the E-step and M-step is
computationally difficult, and many variations of the EM have been
introduced to improve the rate of convergence and/or simplify the
computations. In the case where the E-step cannot be done analytically,
Wei and Tanner~\cite{mcem} suggest replacing the expectation in the
E-step with a Monte Carlo integration. The resulting MCEM algorithm
comes with its own challenges, however; for example, simulating the missing
data $Z_{nj}$, for $j = 1,\ldots,m_n$, from $p(z|\theta_n)$ could be
quite expensive.

Delyon, Lavielle and Moulines \cite{saem} propose, in the case where integration in
the E-step is difficult or intractable, an alternative to the MCEM
using SA.
\begin{saemalg*}
At step $n$, simulate the\break missing data $Z_{nj}$ from the posterior
distribution\break $p(z|  \theta_n)$, $j = 1,\ldots,m_n$. Update
$\widehat{Q}_n(\theta)$ using
\[
\widehat{Q}_n(\theta) = (1-w_n)\widehat{Q}_{n-1}(\theta)
+ \frac{w_n}{m_n} \sum_{j=1}^{m_n} \log f(Z_{nj},\theta),
\]
where $\{w_n\}$ is a sequence as in \eqref{E:sequence}. Then choose
$\theta_{n+1}$ such that $\widehat{Q}_n(\theta_{n+1})
\geq\widehat{Q}_n(\theta)$ for all $\theta\in\Theta$.
\end{saemalg*}

Compared to the MCEM, the SAEM algorithm's use of the simulated data
$Z_{nj}$ is much more efficient. At each iteration, the MCEM simulates
a new set of missing data from the posterior distribution and forgets
the simulated data from the previous iteration. On the other hand, note
that the inclusion of $\widehat{Q}_{n-1}(\theta)$ in the SAEM update
$\theta_n \mapsto\theta_{n+1}$ implies \emph{all} the simulated
data points contribute. It is pointed out in~\cite{saem} that the SAEM
performs strikingly better than the MCEM in problems where maximization
is much cheaper than simulation.

Delyon, Lavielle and Moulines \cite{saem} show, using general SA results, that for a
broad class of complete-data likelihoods $f(z,\theta)$ and under
standard regularity conditions, the SAEM sequence $\{\theta_n\}$
converges a.s. to the set of stationary points $\{\theta\dvtx \nabla
L(\theta) = 0\}$ of the incomplete-data likelihood. Moreover, they
prove that the only attractive stationary points are local maxima; that
is, saddle points of $L(\theta)$ are\break avoided~a.s.

\subsubsection{\texorpdfstring{Adaptive Markov Chain Monte Carlo.}{Adaptive Markov Chain Monte Carlo}}\label{SSS:AM}

A random walk Metropolis (RWM) algorithm is a specific MCMC method that
can be designed to sample from almost any distribution $\pi$. In this
particular case, the proposal is $q(x,y) = q(x-y)$, where $q$ is a
symmetric density. A popular choice of $q$ is a $N_p(0,\Sigma)$
density. It is well known that the convergence properties of Monte
Carlo averages depend on the choice of the proposal covariance matrix
$\Sigma$, in the sense that it affects the rate at which the generated
stochastic process explores the support of $\pi$. Trial and error
methods for choosing $\Sigma$ can be difficult and time consuming. One
possible solution would be to use the history of the process to
suitably tune the proposal. These so-called adaptive algorithms come
with their own difficulties, however. In particular, making use of the
history destroys the Markov property of the process so nonstandard
results are needed in a convergence analysis. For instance, when the
state space contains an atom, Gilks, Roberts and Sahu \cite{gilks} propose an
adaptive algorithm that suitably updates the proposal density only when
the process returns to the atom. The resulting process is not Markov,
but ergodicity is proved using a regeneration argument~\cite{gilks}.

An adaptive Metropolis (AM) algorithm is presented by
Haario, Saksman and Tamminen \cite{haario},
which uses previously visited states to update the
proposal covariance matrix $\Sigma$. Introduce a mean $\mu$ and set
$\theta= (\mu,\Sigma)$. Let $\{w_n\}$ be a deterministic sequence as
in \eqref{E:sequence}.
\begin{amalgg*}
Fix a starting point $z_0$ and initial estimates $\mu_0$ and
$\Sigma_0$. At iteration $n \geq1$ draw $z_n$ from
$N_p(z_{n-1},c\Sigma_{n-1})$ and set
\begin{eqnarray*}
\Sigma_n & =& (1-w_n)\Sigma_{n-1} + w_n(z_n
- \mu_{n-1})(z_n - \mu_{n-1})^\prime,
\\
\mu_n & =& (1-w_n)\mu_{n-1} + w_n z_n.
\end{eqnarray*}
\end{amalgg*}

Note that if $w_n = n^{-1}$, then $\mu_n$ and $\Sigma_n$ are the
sample mean and covariance matrix, respectively, of the observations
$z_1,\ldots,z_n$. The constant $c$ in the AM is fixed and depends only
on the dimension $d$ of the support of $\pi$. A choice of $c$ which
is, in some sense, optimal is $c = 2.4^2/d$ (\cite{casella}, page~316).

It is pointed out in \cite{haario} that the AM has the advantage of
starting the adaptation from the very beginning. This property allows
the AM algorithm to search the support of $\pi$ more effectively
\emph{earlier} than other adaptive algorithms. Note that for the algorithm
of \cite{gilks} mentioned above, the adaptation does not begin until
the atom is first reached; although the renewal times are a.s. finite,
they typically have no finite upper bound.

It is shown in~\cite{haario} that, under certain conditions, the
stationary distribution of the stochastic process $\{z_n\}$ is the
target $\pi$, the chain is ergodic (even though it is no longer
Markovian), and there is almost sure convergence to
$\theta_{\pi} =(\mu_{\pi},\Sigma_{\pi})$, the mean and
covariance of the target
$\pi$. This implies that, as $n \to\infty$, the proposal
distributions in the AM algorithm will be close to the ``optimal''
choice. If $H(z,\theta) = (z-\mu,(z-\mu)(z-\mu)^\prime-\Sigma)$,
then the AM is a general SA algorithm with
$\theta_n = \theta_{n-1} + w_n H(z_n,\theta_{n-1})$, and
Andrieu, Moulines and Priouret \cite{andrieu}
extend the work in~\cite{haario} via new SA
stability results.

\subsubsection{\texorpdfstring{Stochastic approximation Monte Carlo.}{Stochastic approximation Monte Carlo}}\label{SSS:SAMC}

Let $\mathcal{X}$ be a finite or compact space with a dominating
measure $\nu$. Let $p(x) = \kappa p_0(x)$ be a probability density
on $\mathcal{X}$ with
respect to $\nu$ with possibly unknown normalizing constant
$\kappa>0$. We wish to estimate $\int f \,d\nu$, where $f$ is some function
depending on $p$ or $p_0$. For example, suppose $p(x)$ is a prior and
$g(y|x)$ is the conditional density of $y$ given $x$. Then
$f(x) =g(y|x)p(x)$ is the unnormalized posterior density of $x$ and its
integral, the marginal density of $y$, is needed to compute a Bayes factor.

The following stochastic approximation Monte\break Carlo (SAMC) method is
introduced in~\cite{samc}. Let $A_1,\ldots,A_m$ be a
partition of $\mathcal{X}$ and let $\eta_i = \int_{A_i} f \,d\nu$
for $1\leq i \leq m$. Take $\hat{\eta}_i(0)$ as an initial guess, and let
$\hat{\eta}_i(n)$ be the estimate of $\eta_i$ at iteration
$n \geq 1$. For notational convenience, write
\[
\theta_{ni} = \log\hat{\eta}_i(n) \quad\mbox{and} \quad
\theta_n= (\theta_{n1},\ldots,\theta_{nm})^\prime.
\]
The probability vector $\pi= (\pi_1,\ldots,\pi_m)^\prime$ will
denote the \emph{desired} sampling frequency of the $A_i$'s; that is,
$\pi_i$~is the proportion of time we would like the chain to spend in
$A_i$. The choice of $\pi$ is flexible and does not depend on the
particular partition $\{A_1,\ldots,A_m\}$.
\begin{samcalg*}
Starting with initial estimate $\theta_0$, for $n \geq0$ simulate a
sample $z_{n+1}$ using a RWM algorithm with target distribution
%
\begin{equation}\label{E:samc-target}
p(z|\theta_n) \propto\sum_{i=1}^m f(z)e^{-\theta_{ni}} I_{A_i}(z),
\quad z \in\mathcal{X}.
\end{equation}
Then set $\theta_{n+1} = \theta_n + w_{n+1}(\zeta_{n+1} - \pi)$,
where the deterministic sequence $\{w_n\}$ is as in
\eqref{E:sequence}, and $\zeta_{n+1} = (I_{A_1}(z_{n+1}), \ldots,
I_{A_m}(z_{n+1}))^\prime$.
\end{samcalg*}

The normalizing constant in \eqref{E:samc-target} is generally unknown
and difficult to compute. However, $p(z|\theta_n)$ is only used at the
RWM step where it is only required that the target density be known up
to a proportionality constant.

It turns out that, in the case where no $A_i$ are empty, the observed
sampling frequency $\hat{\pi}_i$ of $A_i$ converges to $\pi_i$. This
shows that $\hat{\pi}_i$ is independent of its probability
$\int_{A_i} p\,d\nu$. Consequently,
the resulting chain will not get stuck in
regions of high probability, as a standard Metropolis chain might.

The sequence $\{\theta_n\}$ is a general stochastic approximation and,
using the convergence results of~\cite{andrieu}, Liang, Liu and Carroll
\cite{samc} show that if no $A_i$ is empty and suitable
conditions are met, then
%
\begin{equation}\label{E:samc-limit}
\theta_{ni} \to C + \log\int_{A_i} f \,d\nu- \log\pi_i
\quad \mbox{a.s.,}
\end{equation}
for $1 \leq i \leq m$ as $n \to\infty$, for some arbitrary
constant~$C$. Liang, Liu and Carroll \cite{samc} point out
a \emph{lack of
identifiability} in the limit (\ref{E:samc-limit});
that is, $C$ cannot be determined from $\{\theta_n\}$
alone. Additional information is required, such as
$\sum_{i=1}^m \hat{\eta}_i(n) = c$ for each~$n$
and for some known constant $c$.

\begin{figure}

\includegraphics{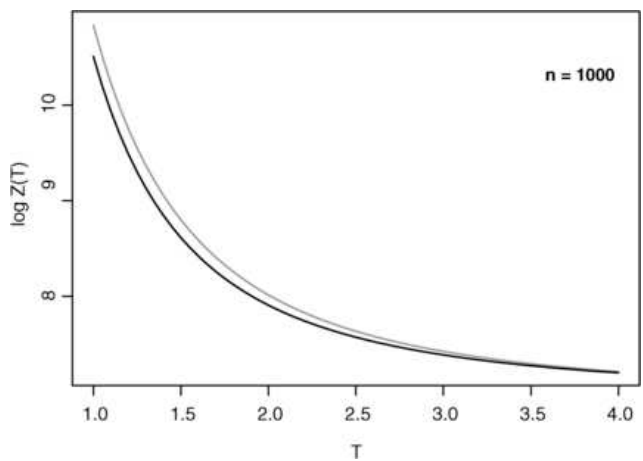}

\caption{$\log Z(T)$ (gray) and SAMC estimate $\log\widehat{Z}(T)$
(black) in Example \protect\ref{EX:samc}.\label{F:samc-ising}}
\end{figure}

In Example~\ref{EX:samc}, we apply SAMC to estimate the partition
function in the one-dimensional Ising model. In this simple situation,
a closed-form expression is available, which we can use as a baseline
for assessing the performance of the SAMC estimate.
\begin{ex}\label{EX:samc}
Consider a one-dimensional Ising model, which assumes that each of the
$d$ particles in a system has positive or negative spin. The Gibbs
distribution on $\mathcal{X}= \{-1,1\}^d$ has density (with respect to counting
measure $\nu$)
\[
p_{T}(x) = \frac{1}{Z(T)} e^{-E(x)/T}, \quad
Z(T) = \sum_{x \in\mathcal{X}}e^{-E(x)/T},
\]
where $T$ is the temperature, and $E$ is\vspace*{2pt} the energy function defined,
in this case, as $E(x) = -\sum_{i=1}^{d-1} x_ix_{i+1}$. The partition
function $Z(T)$ is of particular interest to physicists: the
thermodynamic limit $F(T) = \lim_{d \to\infty} d^{-1}\log Z(T)$ is
used to study phase transitions~\cite{cipra}. In this simple case, a
closed-form expression for $Z(T)$ is available. There are other more
complex systems, however, where no analytical solution is available and
$\nu(\mathcal{X}) = 2^d$ is too large to allow for na\"{\i}ve
calculation of $Z(T)$.

Our jumping-off point is the \emph{energy--temperature duality}
\cite{kzw} $Z(T) = \sum_u \Omega(u) e^{- u/T}$, where
$\Omega(u) = \nu\{x\dvtx E(x)=u\}$ is the density of states.
We will apply SAMC to first
estimate $\Omega(u)$ and then estimate $Z(T)$ with a plug-in:
\[
 \widehat{Z}(T) = \sum_u \widehat{\Omega}(u) e^{-u/T}.
\]
Note here that a single estimate of $\Omega$ can be used to estimate
the partition function for any $T$, eliminating the need for
simulations at multiple temperatures. Furthermore, $\sum_u \Omega(u)
= \nu(\mathcal{X}) = 2^d$ is known so, by imposing this condition on the
estimate $\widehat{\Omega}$ we do not fall victim to the lack of
identifiability mentioned above. Figure~\ref{F:samc-ising} shows the
true partition function $Z(T) = 2^d \cosh^{d-1}(1/T)$ for $d = 10$ as
well as the SAMC estimate $\widehat{Z}(T)$ as a function of
$T \in[1,4]$, on the log-scale, based on $n = 1000$ iterations. Clearly,
$\widehat{Z}$ performs quite well in this example, particularly for
large $T$.
\end{ex}

\subsection{ODE Stability Theory}\label{SS:ode}

The asymptotic theory of ODEs plays an important role in the
convergence analysis of a SA algorithm. After showing the connection
between SA and ODEs, we briefly review some of the ODE theory that is
necessary in the sequel.

Recall the general SA algorithm in \eqref{E:r-m} given by $x_n =
x_{n-1} + w_n y_n$. Assume there is a measurable function $h$ such that
$h(x_{n-1}) = \mathbb{E}[y_n|x_{n-1}]$ and rewrite this algorithm as
\[
x_n = x_{n-1} + w_n h(x_{n-1}) + w_n \{ y_n - h(x_{n-1}) \}.
\]
Define $M_n = y_n - h(x_{n-1})$. Then $\{M_n\}$ is a zero-mean
\textit{martingale} sequence and, under suitable conditions, the martingale
convergence theorem guarantees that $M_n$ becomes negligible as
$n \to\infty$, leaving us with
\begin{eqnarray*}
x_n & =& x_{n-1} + w_n h(x_{n-1}) + w_n M_n
\\
& \approx&  x_{n-1} + w_n h(x_{n-1}).
\end{eqnarray*}
But this latter ``mean trajectory'' is deterministic and essentially a
finite difference equation with small step sizes. Rearranging the terms
gives us
\[
\frac{x_n-x_{n-1}}{w_n} = h(x_{n-1}),
\]
which, for large $n$, can be approximated by the ODE $\dot{x} = h(x)$.
It is for this reason that the study of SA algorithms is related to the
asymptotic properties of solutions to ODEs.

Consider a general autonomous ODE $\dot{x} = h(x)$, where
$h\dvtx \mathbb{R}^d\to\mathbb{R}^d$ is a bounded and
continuous, possibly nonlinear, function.
A solution $x(t)$ of the ODE is a trajectory in $\mathbb{R}^d$ with a given
initial condition $x(0)$. Unfortunately, in many cases, a closed-form
expression for a solution $x(t)$ is not available. For that reason,
other methods are necessary for studying these solutions and, in
particular, their properties as $t \to\infty$.

Imagine a physical system, such as an orbiting celestial body, whose
state is being governed by the ODE $\dot{x} = h(x)$ with initial
condition $x(0) = x_0$. Then, loosely speaking, the system is stable if
choosing an alternative initial condition $x(0) = x_0'$ in a
neighborhood of $x_0$ has little effect on the asymptotic properties of
the resulting solution $x(t)$. The following definition makes this more precise.
\begin{defn}
A point $\xi\in\mathbb{R}^d$ is said to be \emph{locally stable}
for $\dot{x} = h(x)$ if for each $\varepsilon> 0$ there is a $\delta> 0$
such that if
$\|x(0)-\xi\| < \delta$, then $\|x(t)-\xi\| < \varepsilon$ for all
$t \geq 0$. If $\xi$ is locally stable and $x(t) \to\xi$ as $t \to\infty$,
then $\xi$ is \emph{locally asymptotically stable}. If this
convergence holds for all initial conditions $x(0)$, then the
asymptotic stability is said to be \emph{global}.
\end{defn}

Points $\xi$ for which stability is of interest are \emph{equilibrium
points} of $\dot{x} = h(x)$. Any point $\xi$ such that $h(\xi) = 0$
is called an equilibrium point, since the constant solution $x(t)
\equiv\xi$ satisfies $\dot{x} = h(x)$.
\begin{ex}\label{EX:linearstability}
Let $\dot{x} = Ax$, where $A$ is a fixed $d \times d$ matrix. For an
initial condition $x(0) = x_0$, we can write an explicit formula for
the particular solution: $x(t) = e^{At} x_0$ for $t \geq0$. Suppose,
for simplicity, that $A$ has a spectral decomposition $A = U \Lambda
U^\prime$, where $U$ is orthogonal and $\Lambda$ is a diagonal matrix
of the eigenvalues $\lambda_1,\ldots,\lambda_d$ of $A$. Then the
matrix exponential can be written as $e^{At} = U e^{\Lambda t} U^\prime
$, where $e^{\Lambda t}$ is diagonal with $i$th element
$e^{\lambda_i t}$. Clearly, if $\lambda_i < 0$, then
$e^{\lambda_it} \to0$ as $t \to\infty$. Therefore, if $A$ is negative definite,
then the origin $x = 0$ is globally asymptotically stable.
\end{ex}

When explicit solutions are not available, proving asymptotic stability
for a given equilibrium point will require a so-called \emph{Lyapunov
function}~\cite{lasalle}.
\begin{defn}\label{D:lyapunov}
Let $\xi\in\mathbb{R}^d$ be an equilibrium point of the ODE
$\dot{x} = h(x)$ with initial condition $x(0) = x_0$. A
function $\ell\dvtx \mathbb{R}^d \to \mathbb{R}$ is
called a \emph{Lyapunov function} (at $\xi$) if:
\begin{itemize}
\item$\ell$ has continuous first partial derivatives in a
neighborhood of $\xi$;
\item$\ell(x) \geq0$ with equality if and only if $x = \xi$;
\item the time derivative of $\ell$ along the path $x(t)$, defined as
$\dot{\ell}(x) = \nabla\ell(x)^\prime h(x)$, is $\leq0$.
\end{itemize}
A Lyapunov function is said to be \emph{strong} if
$\dot{\ell}(x) =0$ implies $x = \xi$.
\end{defn}

Lyapunov functions are a generalization of the potential energy of a
system, such as a swinging pendulum, and Lyapunov's theory gives a
formal extension of the stability principles of such a system.
Theorem~\ref{T:lyapunov1} is very powerful because it does not require an
explicit formula for the solution. See~\cite{lasalle} for a proof
and various extensions of the Lyapunov theory.
\begin{thm}\label{T:lyapunov1}
If there exists a (strong) Lyapunov function in a neighborhood of an
equilibrium point $\xi$ of $\dot{x} = h(x)$, then $\xi$ is
(asymptotically) stable.
\end{thm}

There is no general recipe for constructing a Lyapunov function. In one
important special case, however, a candidate Lyapunov function is easy
to find. Suppose $h(x) = -\nabla g(x)$, for some positive definite,
sufficiently smooth function $g$. Then $\ell(x) = g(x)$ is a Lyapunov
function since $\dot{\ell}(x) = -\|\nabla g(x)\|^2 \leq0$.
\begin{ex}\label{EX:linearstability2}
Consider again the linear system $\dot{x} = Ax$ from Example
\ref{EX:linearstability}, where $A$ is a $d \times d$ negative definite
matrix. Here we will derive asymptotic stability by finding a Lyapunov
function and applying Theorem~\ref{T:lyapunov1}. In light of the
previous remark, we choose $\ell(x) = -\frac12 x^\prime A x$. Then
$\dot{\ell}(x) = -\|Ax\|^2 \leq0$ so $\ell$ is a strong Lyapunov
function for $\dot{x} = Ax$ and the origin is asymptotically stable by
Theorem~\ref{T:lyapunov1}.
\end{ex}

Of interest is the stronger conclusion of \emph{global}\break asymptotic
stability. Note, however, that Theorem \ref{T:lyapunov1} does not tell
us how far $x_0$ can be from the equilibrium in question and still get
asymptotic stability. For the results that follow, we will prove
the \emph{global} part directly.

\subsection{SA Convergence Theorem}\label{SS:sa-thm}

Consider, for fixed $x_0$ and $\{w_n\}$ satisfying $\eqref
{E:sequence}$, the general SA algorithm
%
\begin{equation}\label{E:gen-sa}
x_n = \operatorname{Proj}_X  \{ x_{n-1} + w_n y_n  \},
\quad n \geq1,
\end{equation}
where $X \subset\mathbb{R}^d$ is compact and
$\operatorname{Proj}_X(x)$ is a projection of
$x$ onto $X$. The projection is necessary when boundedness of the
iterates cannot be established by other means. The \emph{truncated} or
\emph{projected} algorithm \eqref{E:gen-sa} is often written in the
alternative form~\cite{kushner}
%
\begin{equation}\label{E:gen-sa2}
x_n = x_{n-1} + w_ny_n + w_n z_n,
\end{equation}
where $z_n$ is the ``minimum'' $z$ such that $x_{n-1}+w_ny_n+w_n z$
belongs to $X$.

Next we state the main stochastic approximation result used in the
sequel, a special case of Theorem~5.2.3 in~\cite{kushner}.
Define the filtration sequence $\mathscr{F}_n = \sigma(y_1,\ldots,y_n)$.
\begin{thm}\label{T:sa}
For $\{x_n\}$ in \eqref{E:gen-sa} with $\{w_n\}$ satisfying
\eqref{E:sequence}, assume
\begin{enumerate}[$\langle\mathrm{SA1}\rangle$]
\item[$\langle\mathrm{SA1}\rangle$]
$\sup_n \mathbb{E}\|y_n\|^2 <\infty$.
\item[$\langle\mathrm{SA2}\rangle$] There exists a continuous
function $h(\cdot)$ and a random vector $\beta_n$ such that $\mathbb{E}
(y_n|\mathscr{F}_{n-1}) = h(x_{n-1}) + \beta_n$ a.s. for each $n$.
\item[$\langle\mathrm{SA3}\rangle$] $\sum_n w_n \|\beta_n\|$
converges a.s.
\end{enumerate}
If $\xi$ is globally asymptotically stable for $\dot{x} = h(x)$, then
$x_n \to\xi$ a.s.
\end{thm}


\section{Newton's Recursive Estimate}\label{S:mixingestimate}

Let $\Theta$ and $\mathcal{X}$ be the parameter space and sample space,
equipped with $\sigma$-finite measures $\mu$ and $\nu$,
respectively. Typically, $\Theta$ and $\mathcal{X}$ are subsets of Euclidean
space and $\nu$ is Lebesgue or counting measure. The measure $\mu$
varies depending on the inference problem: for estimation, $\mu$ is
usually Lebesgue or counting measure, but for testing, $\mu$ is often
something different (see Example \ref{EX:genes}).

Consider the following model for pairs of random variables
$(X_i,\theta^i) \in\mathcal{X}\times\Theta$:
%
\begin{equation}\label{E:hier-model}
\qquad
\theta^i \stackrel{\mathrm{i.i.d.}}{\sim}f, \quad
X_i|\theta^i\stackrel{\mathrm{ind}}{\sim}p(\cdot|\theta^i), \quad
i=1,\ldots,n,
\end{equation}
where $\{p(\cdot|\theta)\dvtx \theta\in\Theta\}$ is a parametric
family of probability densities with respect to $\nu$ on $\mathcal
{X}$ and $f$
is a probability density with respect to $\mu$ on $\Theta$. In the
present case, the variables (parameters) $\theta^1,\ldots,\theta^n$
are not observed. Therefore, under model \eqref{E:hier-model},\break
$X_1,\ldots,X_n$ are i.i.d. observations from the marginal density
$\Pi_f$ in~\eqref{E:marginal}. We call $f$ the mixing density (or prior,
in the Bayesian context) and the inference problem is to estimate $f$
based on the data observed from $\Pi_f$. The following example gives a
very important special case of this problem---the analysis of DNA
microarray data.
\begin{ex}\label{EX:genes}
A microarray is a tool that gives researchers the ability to
simultaneously investigate the effects of numerous genes on the
occurrence of various diseases. Not all of the genes will be
\textit{expressed}---related to the disease in question---so the problem is to
identify those which are. Let $\theta^i$ represent the expression
level of the $i$th gene, with $\theta^i = 0$ indicating the
gene is not expressed. After some reduction, the data $X_i$ is a
measure of $\theta^i$, and the model is of the form
\eqref{E:hier-model} with $f$ being a prior density with respect to $\mu=
\lambda_{\mathrm{Leb}} + \delta_{\{0\}}$. Consider the multiple
testing problem
\[
H_{0i}\dvtx \theta^i = 0, \quad i=1,\ldots,n.
\]
The number $n$ of genes under investigation is often in the thousands
so, with little information about $\theta^i$ in $X_i$, choosing a
fixed prior $f$ would be problematic. On the other hand, the data
contain considerable information about the prior $f$ so the
\textit{empirical Bayes} approach---using the data to \emph{estimate} the
prior---has been quite successful~\cite{efron2001}.
\end{ex}

In what follows, we focus our attention on a particular estimate of the
mixing density $f$. Let $x_1,\ldots, x_n \in\mathcal{X}$ be i.i.d. observations
from the mixture density $\Pi_f$ in \eqref{E:marginal}.
Newton~\cite{newton} suggests the following algorithm for estimating $f$.
\begin{newtonalg*}
Choose a positive density $f_0$ on $\Theta$ and weights $w_1,\ldots
,w_n \in(0,1)$. Then for $i=1,\ldots,n$, compute
%
\begin{equation}\label{E:newton1}
\qquad
f_i(\theta) = (1-w_i)f_{i-1}(\theta) + w_i
\frac{p(X_i|\theta)f_{i-1}(\theta)}{\Pi_{i-1}(X_i)},\hspace*{-8pt}
\end{equation}
where $\Pi_j(x) = \int p(x|\theta) f_j(\theta)\,d\mu(\theta)$, and
report $f_n(\theta)$ as the final estimate.
\end{newtonalg*}

In the following subsections we establish some\break asymptotic properties of
$f_n$ as $n \to\infty$ and we show the results of several numerical
experiments that demonstrate the finite-sample accuracy of Newton's
estimate~\eqref{E:newton1} in both the discrete and continuous cases.
First, a few important remarks.
\begin{itemize}
\item The update $f_{i-1} \mapsto f_i$ in \eqref{E:newton1} is similar
to a Bayes estimate based on a Dirichlet process prior (DPP), given the
information up to, and including, time $i-1$. That is, after observing
$X_1,X_2,\ldots,X_{i-1}$, a~Bayesian might model $f$ with a DPP
$\mathscr{D}(\frac{1-w_i}{w_i},\break  f_{i-1})$. In this case, the posterior
expectation is exactly the $f_i$ in \eqref{E:newton1}.
\item Because $f_n$ depends on the ordering of the data and not simply
on the sufficient statistic $(X_{(1)},\ldots,\break X_{(n)})$, it is
\textit{not} a posterior quantity.
\item The algorithm is very fast: if one evaluates \eqref{E:newton1}
on a grid of $m$ points $\theta_1,\ldots,\theta_m$ and calculates
the integral in $\Pi_{i-1}$ using, say, a trapezoid rule, then the
computational complexity is $mn$.
\end{itemize}

\subsection{Review of Convergence Results}\label{SS:newton-review}

In this section, we give a brief review of the known convergence
results for Newton's estimate $f_n$ in the case of a finite parameter
space. The case of a compact~$\Theta$ is quite different and, until
very recently~\cite{tmg}, nothing was known about the convergence of
$f_n$ in such problems; see Section~\ref{S:discuss}.

Newton~\cite{newton}, building on the work in~\cite{nqz,newtonzhang},
states the following convergence theorem.
\begin{thm}\label{T:newton}
Assume the following:
\begin{enumerate}[$\langle\mathrm{N2}\rangle$]
\item[$\langle\mathrm{N1}\rangle$] $\Theta$ is finite and $\mu$
is counting measure.
\item[$\langle\mathrm{N2}\rangle$] $\sum_n w_n = \infty$.
\item[$\langle\mathrm{N3}\rangle$] $p(x|\theta)$ is bounded away
from $0$ and $\infty$.
\end{enumerate}
Then \emph{surely} there exists a density $f_{\infty}$ on $\Theta$
such that $f_n \to f_{\infty}$ as $n \to\infty$.
\end{thm}

Newton~\cite{newton} presents a proof of Theorem~\ref{T:newton} based
on the theory of nonhomogeneous Markov chains. He proves that $f_n$
represents the $n$-step marginal distribution of the Markov chain $\{
Z_n\}$ given by
\[
Z_0 \sim f_0, \quad Z_n = \cases{%
Z_{n-1}, & with prob $1-w_n$, \cr
Y_n, & with prob $w_n$,}
\]
where $Y_n$ has density $\propto p(X_n|\theta)f_{n-1}(\theta)$.
However, the claim that this Markov chain admits a stationary
distribution is incomplete---N2 implies the chain $\{Z_n\}$ is weakly
ergodic but the necessary strong ergodicity property does not follow,
even when $\Theta$ is finite. Counterexamples are given in
\cite{isaacson,ghosh}.
Ghosh and Tokdar~\cite{ghosh} prove consistency of $f_n$ along quite
different lines. For probability densities $\psi$ and $\varphi$ with
respect to~$\mu$, define the Kullback--Liebler (KL) divergence,
%
\begin{equation}\label{E:kl}
K(\psi,\varphi) = \int_{\Theta} \psi \log(\psi/\varphi) \,d\mu.
\end{equation}
The following theorem is proved in~\cite{ghosh} using an approximate
martingale representation of $K(f,f_n)$.
\begin{thm}\label{T:ghosh-tokdar}
In addition to \emph{N1}--\emph{N3}, assume
\vspace{-1mm}
\begin{enumerate}[$\langle\mathrm{GT2}\rangle$]
\item[$\langle\mathrm{GT1}\rangle$] $\sum_n w_n^2 < \infty$.
\item[$\langle\mathrm{GT2}\rangle$] $f$ is \textup{identifiable};
that is, $f \mapsto\Pi_f$ is \textit{injective}.
\end{enumerate}
Then $K(f,f_n) \to0$ a.s. as $n \to\infty$.
\end{thm}

Part of the motivation for the use of the KL divergence lies in the
fact that the ratio $f_n/f_{n-1}$ has a relatively simple form. More
important, however, is the Lyapunov property shown in the proof Theorem
\ref{T:martin-ghosh}. Sufficient conditions for GT2 in the case of
finite $\Theta$ are given in, for example, \cite{teicher,lindsay}. San
Martin and Quintana~\cite{quintana2002} also discuss the issue of
identifiability in connection with the consistency of $f_n$.

\subsection{Newton's Estimate as a SA}\label{SS:newtonconvsa}

Here we show that Newton's algorithm \eqref{E:newton1} is a special
case of SA. First, note that if $f$ is viewed as a prior density, then
estimating $f$ is an \emph{empirical Bayes} (EB) problem. The ratio in
\eqref{E:newton1} is nothing but the posterior distribution of
$\theta$, given $x_i$, and assuming that the prior~$f$
is equal to $f_{i-1}$.
This, in fact, is exactly the approach taken in Example~\ref{EX:eb} to
apply SA in an EB problem.

Let $\mu$ be counting measure and $d = \mu(\Theta)$. We can think of
$f_n(\theta)$ as a vector $f_n = (f_n^1,\ldots,f_n^d)^\prime$ in the
probability simplex $\Delta^d$, defined as
\[
\Delta^d =  \Biggl\{(\varphi^1,\ldots,\varphi^d)' \in[0,1]^d\dvtx
\sum_{i=1}^d \varphi^i = 1 \Biggr\}.
\]
Define $H\dvtx \mathcal{X}\times\Delta^d \to\mathbb{R}^d$ with
$k$th component
%
\begin{equation}\label{E:H-function1}
\qquad
H_k(x,\varphi) = \frac{p(x|\theta_k) \varphi^k}{\Pi_{\varphi}(x)}
- \varphi^k, \quad k=1,\ldots,d,\hspace*{-5pt}
\end{equation}
where $\Pi_\varphi(x) = \sum_k p(x|\theta_k)\varphi^k$ is the marginal
density on $\mathcal{X}$ induced by $\varphi\in\Delta^d$. Then
\eqref{E:newton1} becomes
%
\begin{equation}\label{E:newton2}
f_n = f_{n-1} + w_n H(X_n,f_{n-1}).
\end{equation}
Let $P_x = \operatorname{diag}\{p(x|\theta_k)\dvtx k=1,\ldots,d\}$ be the
diagonal matrix of the sampling density values and define the mapping
$h: \Delta^d \to\mathbb{R}^d$ to be the conditional expectation of
$H(x,f_n)$, given $f_n = \varphi$:
%
\begin{eqnarray}\label{E:h-function}
h(\varphi) & =& \int_{\mathcal{X}} H(x,\varphi) \Pi_f(x) \,d\nu(x)
\nonumber\\[-8pt]
\\[-8pt]
& =& \int_{\mathcal{X}} \frac{\Pi_f(x)}{\Pi_{\varphi}(x)}
P_x \varphi\,d\nu(x) - \varphi,
\nonumber
\end{eqnarray}
where $f = (f^1,\ldots,f^d)^\prime$ is the true mixing/prior
distribution. From \eqref{E:h-function}, it is clear that $f$ solves
the equation $h(\varphi) = 0$ which implies (i) $f$ is an equilibrium
point of the ODE $\dot{\varphi} = h(\varphi)$, and (ii) that $f$ is a
fixed point of the map
\[
T(\varphi) = h(\varphi) + \varphi= \int\frac{\Pi_f(x)}
{\Pi_\varphi(x)} P_x \varphi\,d\nu(x).
\]
Newton~\cite{newton}, page 313, recognized the importance of this map in
relation to the limit of $f_n$. Also, the use of $T$ in~\cite
{csiszar,shyamalkumar} for the $I$-projection problem is closely
related to the SA approach taken here.

We have shown that \eqref{E:newton2} can be considered as a general SA
algorithm, targeting the solution $\varphi=f$ of the equation
$h(\varphi) =0$ in $\Delta^d$. Therefore, the SA results of
Section~\ref{SS:sa-thm} can be used in the convergence analysis. The following
theorem is proved in Appendix~\ref{SS:app-2}.\looseness=1
\begin{thm}\label{T:martin-ghosh}
Assume \textup{N1}, \textup{N2}, \textup{GT1} and \textup{GT2}.
If $p(\cdot|\theta) > 0$ $\nu$-a.e. for each $\theta$, then $f_n \to f$
a.s.
\end{thm}
\begin{remark}\label{RE:extension1}
Removal of the boundedness condition N3 on $p(x|\theta)$ in
Theorem~\ref{T:martin-ghosh} extends the consistency result of~\cite
{ghosh} to many important cases, such as mixtures of normal or gamma densities.
\end{remark}
\begin{remark}\label{RE:boundary}
Theorem~\ref{T:martin-ghosh} covers the \emph{interior} case (when
$f$ is strictly positive) as well as the \emph{boundary} case (when
$f^i = 0$ for some $i$). The fact that $f_0^i > 0$ implies $f_n^i > 0$
for all $n$ suggests that convergence may be slow in the boundary case.
\end{remark}

\subsection{Simulations}\label{SS:examples}

\begin{figure*}

\includegraphics{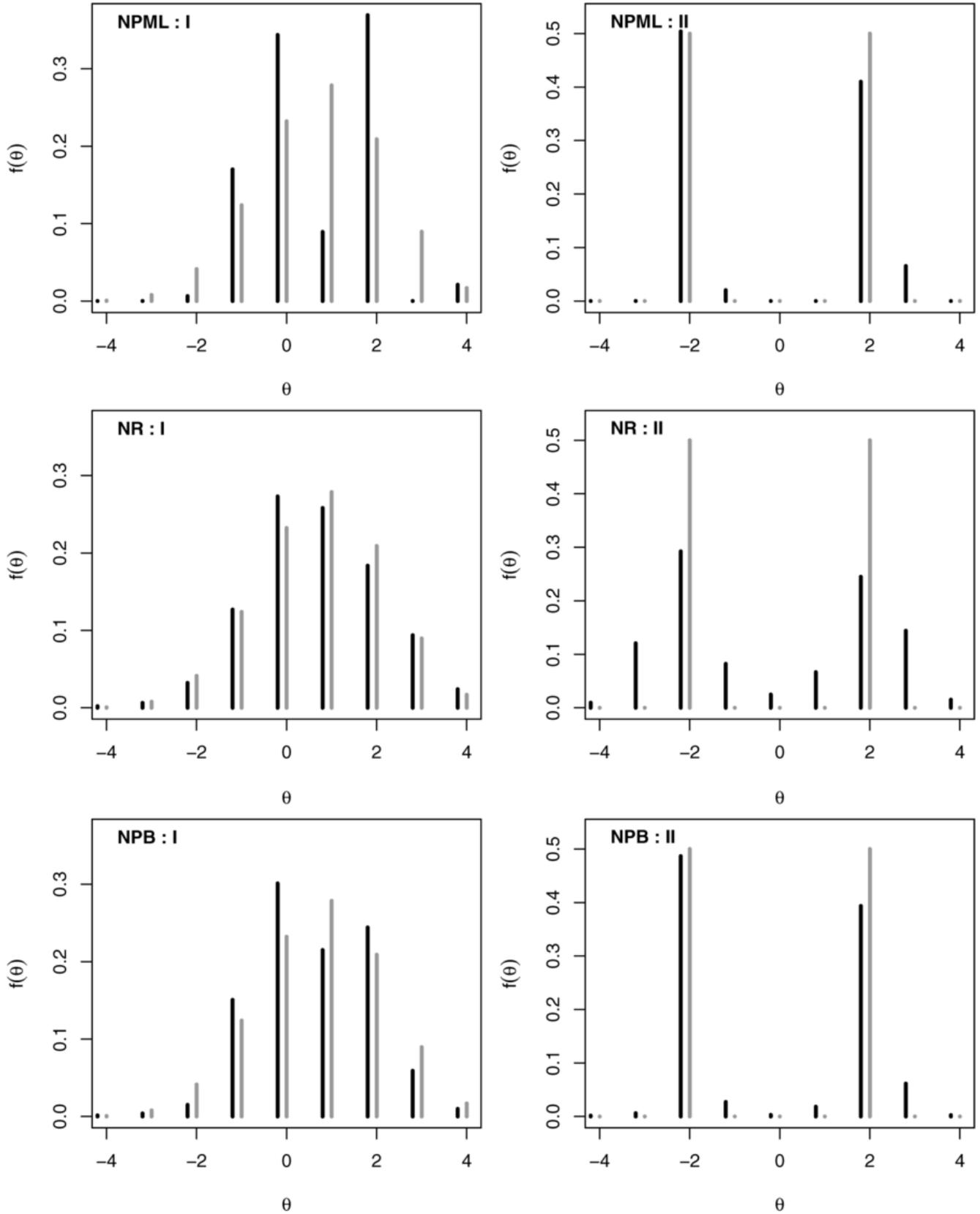}

\caption{Estimates of mixing densities \textup{I} and \textup{II} in Example
\protect\ref{EX:newton-discrete}. Left column: True $f$ (gray) for model
\textup{I} and the three estimates (black). Right column:
True $f$ (gray) for model \textup{II} and the three estimates
(black).\label{F:n-discrete}}
\end{figure*}

Here we provide numerical illustrations comparing the performance of
Newton's estimate with that of its competitors. We consider a
location-mixture of normals; that is, $p(\cdot|\theta)$ is a
$N(\theta,\sigma^2)$ density. The weights are set to be $w_i =
(i+1)^{-1}$ and the initial estimate $f_0$ is taken to be a
$\operatorname{Unif}(\Theta)$ density. For the Bayes estimate, we assume a Dirichlet
process prior $f \sim\mathscr{D}(1,f_0)$ in each example.
\begin{ex}[(\textit{Finite $\Theta$})]\label{EX:newton-discrete}
In this example, we compare Newton's recursive (NR) estimate with the
nonparametric maximum likelihood (NPML) estimate and the nonparametric
Bayes (NPB) estimate. Computation of NR and NPML (using the EM
algorithm) is straightforward. Here, in the case of finite $\Theta$,
we use sequential imputation \cite{liu} to calculate NPB.
Take $\Theta= \mathbb{Z}\cap[-4,4]$, and set $\sigma= 1$ in
$p(x|\theta)$.
We consider two different mixing distributions on $\Theta$:
\begin{enumerate}[II.]
\item[I.] $f = \operatorname{Bin}(8,0.6)$,
\item[II.] $f = 0.5\delta_{\{-2\}} + 0.5 \delta_{\{2\}}$.
\end{enumerate}
We simulate 50 data sets of size $n=100$ from the models corresponding
to mixing densities I, II and computing the three estimates for each.
Figure~\ref{F:n-discrete} shows the resulting estimates for a randomly
chosen data set from each model. Notice that NR does better for model I
than both NPML and NPB. The story is different for model II---both NPML
and NPB are considerably better than NR. This is further illustrated in
Figure~\ref{F:kl-boxplot} where the KL divergence $K(\Pi_f,\widehat
{\Pi}_n)$ on $\mathcal{X}= \mathbb{R}$ is summarized over the 50
samples. We see that
NR has a slightly smaller KL number than NPML and NPB for model I, but
they clearly dominate NR for model II. This discrepancy is at least
partially explained by Remark~\ref{RE:boundary}; see Section~\ref
{S:discuss} for further discussion. We should point out, however, that
both NPML and NPB take significantly longer to compute than NR, about
100 times longer on average.
\end{ex}

%
\begin{ex}[(\textit{Compact $\Theta$})]\label{EX:newton-compact}
We consider a one- and a two-component mixture of beta densities on
$\Theta= [0,1]$ as the true $f$:
\begin{enumerate}[II.]
\item[I.] $f = \operatorname{Beta}(2,7)$,
\item[II.] $f = 0.33 \operatorname{Beta}(3,30) + 0.67\operatorname{Beta}(4,4)$.
\end{enumerate}
Let $\sigma= 0.1$ be the normal sampling variance. Again, computation
of NR is straightforward. To compute NPB, the importance sampling
algorithm in~\cite{tmg} that makes use of a collapsing of the
Poly\'{a} Urn scheme is used. Figure~\ref{F:n-compact} shows a typical
realization of NR and NPB, based on a sample of size $n = 100$ from
each of the corresponding marginals. Note that the Bayes estimate does
a rather poor job here, being much too spiky in both cases. This is
mainly because the posterior for $f$ sits on discrete distributions. On
the other hand, Newton's estimate has learned the general shape of $f$
after only 100 iterations and results in a much better estimate than
NPB. Furthermore, on average, the computation time for NR is again
about 100 times less than that of NPB.
\end{ex}

\setcounter{figure}{5}
\begin{figure*}[b]

\includegraphics{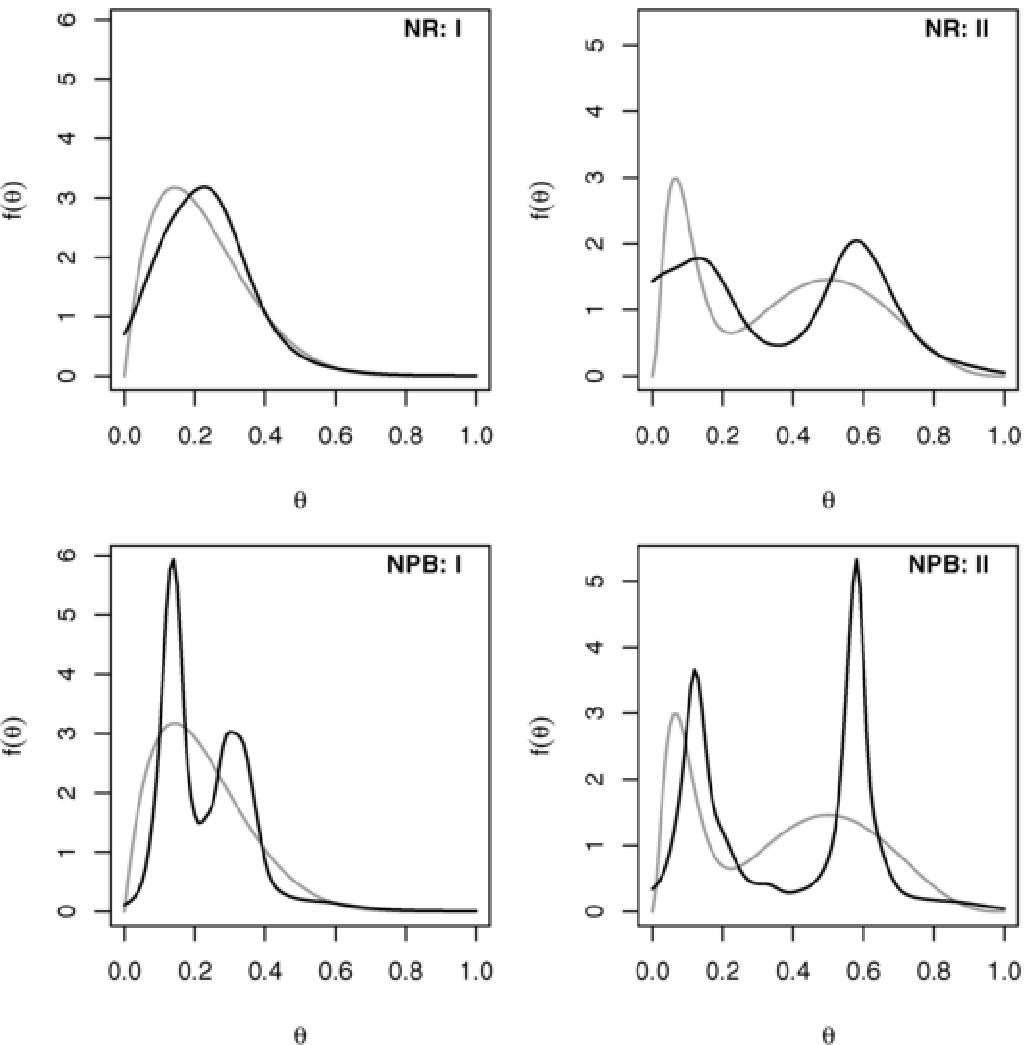}

\caption{Estimates of the mixing densities \textup{I} and \textup{II} in Example
\protect\ref{EX:newton-compact}. Top row: true $f$ (gray)
and \emph{NR} (black). Bottom row: true $f$ (gray) and \emph{NPB}
(black).\label{F:n-compact}}
\end{figure*}

\section{N$+$P Algorithm}\label{S:NplusP}

\setcounter{figure}{4}
\begin{figure}

\includegraphics{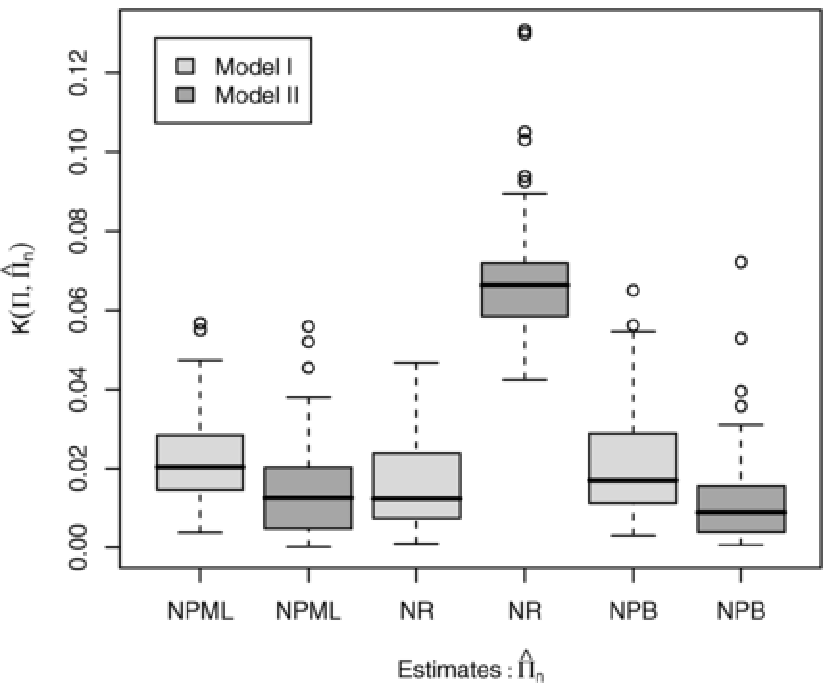}

\caption{Summary of the KL divergence $K(\Pi_f,\widehat{\Pi}_n)$
for the three estimates $\widehat{\Pi}_n$ in models \textup{I} and \textup{II} in
Example \protect\ref{EX:newton-discrete}.\label{F:kl-boxplot}}
\end{figure}

Suppose that the sampling distribution on $\mathcal{X}$ is
parametrized not
only by $\theta$ but by an additional parameter~$\xi$. An example of
this is the normal distribution with mean $\theta$ and variance
$\xi=\sigma^2$. More specifically, we replace the sampling densities
$p(x|\theta)$ of Section~\ref{S:mixingestimate} with $p(x|\theta,\xi)$
where $\theta$ is the latent variable, and $\xi$ is also unknown.
Newton's algorithm cannot be used in this situation since $\theta$
does not fully specify the sampling density.

In this section we introduce a modification of Newton's algorithm to 
simultaneously and recursively estimate both a mixing distribution
and an additional unknown parameter. This modification, called the
Newton$+$Plug-in (N$+$P), is actually quite\vadjust{\goodbreak} simple---at each step we use a
plug-in estimate of $\xi$ in the update~\eqref{E:newton1}. We show
that the N$+$P algorithm can be written as a general SA algorithm and,
under certain conditions, prove its consistency.

Let $p(x|\theta,\xi)$ be a two-parameter family of densities on
$\mathcal{X}$, and consider the model
%
\begin{eqnarray}\label{E:REmodel}
\theta^1,\ldots,\theta^n & \stackrel{\mathrm{i.i.d.}}{\sim} & f,
\nonumber\\[-8pt]
\\[-8pt]
\qquad
X_{i1},\ldots,X_{ir} & \stackrel{\mathrm{i.i.d.}}{\sim}&
p(\cdot |\theta^i,\xi), \quad i=1,\ldots,n,
\nonumber
\end{eqnarray}
where $f$ is an unknown density on $\Theta$ and the parameter
$\xi\in\Xi$ is also unknown. The number of replicates $r \geq2$ is assumed
fixed. Note that \eqref{E:REmodel} is 
simply a \textit{nonparametric} random effects model.

Assume, for simplicity, that $\Xi\subseteq\mathbb{R}$; the more
general case
$\Xi\subseteq\mathbb{R}^p$ is a natural extension of what follows. Let
$\Theta= \{\theta_1,\ldots,\theta_d\}$ be a \emph{finite} set and
take $\mu$ to be counting measure on $\Theta$. Recall that
$\Delta^d$ is the probability simplex. Assume:
\begin{enumerate}[$\langle\mbox{NP2}\rangle$]
\item[$\langle\operatorname{NP1}\rangle$] $\xi\in\mbox{int}(\Xi_0)$,
where $\Xi_0$ is a compact and convex subset of $\Xi$.
\item[$\langle\operatorname{NP2}\rangle$] $f \in\mbox{int}(\Delta_0)$
where $\Delta_0 \subset\Delta^d$ is compact and, for each
$\varphi\in\Delta_0$, the coordinates $\varphi^1,\ldots,\varphi^d$
are bounded away from zero.
\end{enumerate}
The subset $\Xi_0$ can be arbitrarily large so assumption NP1 causes
no difficulty in practice. Assumption NP2 is somewhat restrictive in
that $f$ must be strictly positive. While NP2 seems necessary to prove
consistency (see Appendix \ref{SS:app-3}), simulations suggest that
this assumption can be weakened.

The N$+$P algorithm uses an estimate of $\xi$ at each step in Newton's
algorithm. We assume here that an \emph{unbiased} estimate is available:
\begin{enumerate}[$\langle\mbox{NP3}\rangle$]
\item[$\langle\mbox{NP3}\rangle$] There exists an unbiased estimate
$T_{\mbox{{\tiny UBE}}}(x)$, $x \in\mathcal{X}^r$, of $\xi$ with
variance $v^2 <\infty$.
\end{enumerate}
Later we will replace the unbiased estimate with a Bayes estimate. This
will require replacing NP3 with another assumption.

At time $i=1,\ldots,n$, we observe an $r$-vector $X_i = (X_{i1},\ldots
,X_{ir})^\prime$ and we compute $\hat{\xi}^{(i)} = T_{\mbox{{\tiny UBE}}}(X_i)$.
An unbiased estimate of $\xi$ based on the entire data
$X_1,\ldots,\break  X_n$ would be the average $\xi_n = n^{-1}\sum_{i=1}^n
\hat{\xi}^{(i)}$, which has a convenient recursive expression
%
\begin{equation}\label{E:xi-n}
\qquad
\xi_i = i^{-1}  \bigl[ (i-1)\xi_{i-1} + \hat{\xi}^{(i)}  \bigr],
\quad i=1,\ldots,n.
\end{equation}
More importantly, by construction,
$\hat{\xi}^{(1)},\ldots,\hat{\xi}^{(n)}$ are i.i.d.
random variables with mean $\xi$ and finite variance.
It is, therefore, a consequence of the SLLN that $\xi_n$, as defined
in \eqref{E:xi-n}, converges a.s. to $\xi$. While this result holds
for any unbiased estimate $T$, an unbiased estimate $T'$ with smaller
variance is preferred, since it will have better finite-sample performance.

Define the mapping $H\dvtx \mathcal{X}^r \times\Delta_0 \times\Xi_0
\to\mathbb{R}^d$ with $k$th component
%
\begin{equation}\label{E:H-function2}
H_k(x,\varphi,\psi) = \frac{\overline{p}(x|\theta_k,\psi)
\varphi^k}{\sum_j\overline{p}(x|\theta_j,\psi) \varphi^j} - \varphi^k,
\end{equation}
for $k=1,\ldots,d$, where $\varphi$ and $\psi$ denote generic elements
in $\Delta_0$ and $\Xi_0$, respectively, and
$\overline{p}(\cdot|\theta,\psi)$ is the
joint density of an i.i.d. sample of size $r$ from
$p(\cdot|\theta,\psi)$.
\begin{NPalg*}
Choose an initial estimate $f_0 \in\Delta_0$, weights
$w_1,\ldots,w_n \in(0,1)$, and an arbitrary
$\xi_0 \in\Xi_0$. Then for $i=1,\ldots,n$ compute
\begin{eqnarray*}
\xi_i & = & \operatorname{Proj}_{\Xi_0}
\bigl\{ i^{-1} \bigl[ (i-1)\xi_{i-1} + \hat{\xi}^{(i)}  \bigr]  \bigr\} ,
\\
f_i & =& \operatorname{Proj}_{\Delta_0}  \bigl\{ f_{i-1} + w_i
H(X_i;f_{i-1},\xi_i) \bigr\},
\end{eqnarray*}
and produce $(f_n,\xi_n)$ as the final estimate.
\end{NPalg*}

We claim that the N$+$P algorithm for estimating $f$ can be written as a
general SA involving the true but unknown $\xi$ plus an additional
perturbation. Define the quantities
%
\begin{eqnarray}
\quad
h(f_{n-1}) & =& \mathbb{E}[H(X_n,f_{n-1},\xi)|\mathscr{F}_{n-1}],
\label{E:np-h} \\
\beta_n & =& \mathbb{E}[H(X_n,f_{n-1},\xi_n)|\mathscr{F}_{n-1}]
\nonumber\\[-8pt]\label{E:beta1}
\\[-8pt]
&&{} - \mathbb{E}[H(X_n,f_{n-1},\xi)|\mathscr{F}_{n-1}] ,
\nonumber
\end{eqnarray}
where $\mathscr{F}_{n-1} = \sigma(X_1,\ldots,X_{n-1})$, so that
\[
\mathbb{E}[H(X_n,f_{n-1},\xi_n)|\mathscr{F}_{n-1}]
= h(f_{n-1}) + \beta_n.
\]
Now the update $f_{n-1} \mapsto f_n$ can be written as
%
\begin{equation}\label{E:NP2}
\qquad
f_n = f_{n-1} + w_n \{h(f_{n-1}) + M_n + \beta_n + z_n\},
\end{equation}
where $z_n$ is the ``minimum'' $z$ keeping $f_n$ in $\Delta_0$, and
\[
M_n = H(X_n,f_{n-1},\xi_n) - h(f_{n-1})-\beta_n
\]
is a martingale adapted to $\mathscr{F}_{n-1}$. Notice that
\eqref{E:NP2} is now in a form in which
Theorem \ref{T:sa} can be applied. We will make
use of the Law of Iterated Logarithm so define $u(t) = (2t\log\log
t)^{1/2}$. The consistency properties of the N$+$P algorithm are
summarized in the following theorem.
\begin{thm}\label{T:martin-ghosh2}
Assume \textup{N1}, \textup{GT1}, \textup{GT2}, \textup{NP1--NP3}.
In addition, assume
\begin{enumerate}[$\langle\mathrm{NP4}\rangle$]
\item[$\langle\mathrm{NP4}\rangle$] $\frac{\partial}{\partial\psi}
H(x;\varphi,\psi)$ is bounded on $\mathcal{X}^r
\times\Delta_0\times\Xi_0$.
\item[$\langle\mathrm{NP5}\rangle$] $\sum_n w_n n^{-1} u(n)$ converges.
\end{enumerate}
Then $(f_n,\xi_n) \to(f,\xi)$ a.s. as $n \to\infty$.
\end{thm}

We now remove the restriction to unbiased estimates of $\xi$, focusing
primarily on the use of a Bayes estimate in place of the unbiased
estimate. But first, let $\tilde{\xi}_i = T(X_1,\ldots,X_i)$ be any
suitable estimate of $\xi$ based on only $X_1,\ldots,X_i$. Then
replace the N$+$P update $f_{i-1} \mapsto f_i$ with
\[
\tilde{f}_i = \operatorname{Proj}_{\Delta_0}
\{ \tilde{f}_{i-1} + w_iH(X_i,\tilde{f}_{i-1},\tilde{\xi}_i)  \}.
\]
While this adaptation is more flexible with regard to the choice of
estimate, this additional flexibility does not come for free. Notice
that the algorithm is no longer \emph{recursive}. That is, given a new
data point $x_{n+1}$, we need more information than just the pair
$(\tilde{f}_n,\tilde{\xi}_n)$ to obtain
$(\tilde{f}_{n+1},\tilde{\xi}_{n+1})$.
\begin{cor}\label{cor:cor1}
If assumptions \textup{NP3} and \textup{NP5}
in Theorem~\ref{T:martin-ghosh2} are replaced by
\begin{enumerate}[$\langle\mbox{NP5}' \rangle$]
\item[$\langle\mbox{NP3}' \rangle$]
$|\tilde{\xi}_n-\xi| = O(\rho_n)$ a.s. as $n \to\infty$,
\item[$\langle\mbox{NP5}' \rangle$] $\sum_n w_n \rho_n <\infty$,
\end{enumerate}
then $(\tilde{f}_n,\tilde{\xi}_n) \to(f,\xi)$ a.s. as
$n\to \infty$.
\end{cor}

Typically, for Bayes and ML estimates, the rate is
$\rho_n =n^{-1/2}$. Then NP5$'$ holds if,
e.g., $w_n \sim n^{-1}$.

To illustrate the N$+$P and its modified version, consider the special
case where $p(\cdot|\theta,\xi)$ in \eqref{E:REmodel} is a normal
density with mean $\theta$ and $\xi= \sigma^2$ is the unknown
variance. That is,
\[
X_{i1},\ldots,X_{ir} \stackrel{\mathrm{i.i.d.}}{\sim}
N(\theta^i,\sigma^2), \quad i=1,\ldots,n.
\]
Moreover, the statistic $S_i = \overline{X}_i$ is sufficient for the
mean and
the density $g(\cdot|\theta,\sigma^2)$ of $S_i$ is known. Therefore,
$H$ in \eqref{E:H-function2} can be written as
%
\begin{equation}\label{E:H-function3}
H_k(s,\varphi,\psi) = \frac{g(s|\theta_k,\psi) \varphi^k}{\sum_j
g(s|\theta_j,\psi) \varphi^j} - \varphi^k
\end{equation}
for $k=1,\ldots,d$, where $g(s|\theta,\psi)$ is the $N(\theta,\psi
/r)$ density. Even in this simple example, it is not obvious that the
function $H$ in \eqref{E:H-function3} satisfies NP4. A proof of the
following proposition is in Appendix~\ref{SS:app-3}.
\begin{prop}\label{P:normal-mix}
\textup{NP4} holds for $H$ in \eqref{E:H-function3}.
\end{prop}

Let $\Sigma_0$ be the $\Xi_0$ defined in the general setup. For the
N$+$P, we choose $T_{\mbox{{\tiny UBE}}}(x)$ to be the sample variance of
$x$, resulting in the recursive estimate
%
\begin{equation}\label{E:unbiased}
\sigma_i^2 = \frac{1}{i(r-1)} \sum_{k=1}^i \sum_{j=1}^r
(X_{kj}-\overline{X}_k)^2.
\end{equation}
For $\sigma^2$, take the standard noninformative prior
$p(\sigma^2)= (\sigma^2)^{-1}$. Under squared-error
loss, the Bayes estimate of
$\sigma^2$ based on $X_1,\ldots,X_i$ is
%
\begin{eqnarray}\label{E:bayes}
\tilde{\sigma}_i^2 & =& \mathbb{E}(\sigma^2 | \mathscr{F}_i)
\nonumber\\[-8pt]
\\[-8pt]
& =& \frac{1}{i(r-1)-2} \sum_{k=1}^i \sum_{j=1}^r
(X_{kj}-\overline{X}_k)^2.
\nonumber
\end{eqnarray}
Note that $|\tilde{\sigma}_n^2-\sigma^2| = O(n^{-1/2})$ a.s. so the
conclusion of Corollary \ref{cor:cor1} holds if $w_n \sim n^{-1}$.

The following example compares three resulting estimates for this
location mixture of normals problem: when $\sigma^2$ is known, when
\eqref{E:unbiased} is used with the N$+$P and when \eqref{E:bayes} is
used in the modified N$+$P. Convergence of the iterates holds in each
case by Theorems \ref{T:martin-ghosh} and \ref{T:martin-ghosh2} and
Corollary \ref{cor:cor1}.
\begin{ex}\label{EX:NP-example}
Let $\Theta= \mathbb{Z}\cap[-4,4]$ and take $f$ to
be a $\operatorname{Bin}(8,0.5)$
density on $\Theta$. Suppose $r = 10$, $n = 100$, $w_i = (i+1)^{-1}$
and set $\sigma^2 = 1.5$. For each of 100 simulated data sets, the
three estimates of $f$ are computed using Newton's algorithm, the N$+$P
and the Bayes modification. Each algorithm produces estimates
$\hat{f}$ and $\hat{\sigma}^2$ with which we compute
$\widehat{\Pi}_{\hat{\sigma}}(s) = \sum_{j=1}^d g(s|\theta_j,\hat{\sigma}^2)
\hat{f}^j$. Figure \ref{F:NP-example} summarizes the 100 KL
divergences $K(\Pi,\widehat{\Pi}_{\hat{\sigma}})$ for each of the
three estimates. Surprisingly, little efficiency is lost when an
estimate of $\sigma^2$ is used rather than the true value. Also, the
N$+$P and the Bayes modification perform comparably, with the Bayes
version performing perhaps slightly better on average. Note that no
projections onto $\Sigma_0 = [10^{-4},10^4]$ or
\[
\Delta_0 = \{\varphi\in\Delta\dvtx \varphi^k \geq10^{-4}, k=1,\ldots,d \}
\]
were necessary in this example.
\end{ex}

\section{Discussion}\label{S:discuss}

\setcounter{figure}{6}
\begin{figure}

\includegraphics{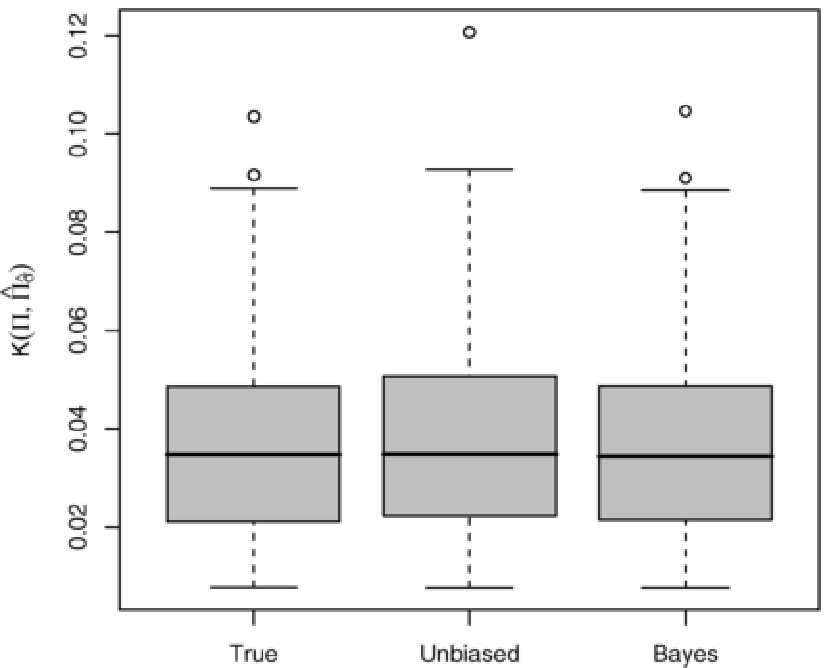}

\caption{Summary of KL divergences
$K(\Pi,\widehat{\Pi}_{\hat{\sigma}})$ for the three algorithms in
Example \protect\ref{EX:NP-example}.\label{F:NP-example}}
\end{figure}

In this paper, we have used general results in the area of stochastic
approximation to prove a consistency theorem for a recursive estimate
of a mixing distribution/prior in the case of a finite parameter space
$\Theta$. It is natural to wonder if this theorem can be extended to
the case where $f$ is an infinite-dimensional parameter on an
uncountable space $\Theta$. Very recently, Tokdar, Martin and Ghosh \cite{tmg}
have proved consistency of $f_n$ in the infinite-dimensional case,
under mild conditions. Their argument is based on the approximate
martingale representation used in~\cite{ghosh} but applied to the KL
divergence $K(\Pi_f,\Pi_n)$ between the induced marginals. Again,
there is a connection between their approach and the SA approach taken
here, namely, $K(\Pi_f,\Pi_\varphi)$ is also a Lyapunov function for
the associated ODE $\dot{\varphi} = h(\varphi)$.

In addition to convergence, there are some other interesting
theoretical and practical questions to consider. First and foremost,
there is the question of \emph{rate of convergence} which, from a
practical point of view, is much more important than convergence alone.
We expect that, in general, the rate of convergence will depend on the
support of $f_0$, the weights $w_n$ and, in the case of an uncountable
$\Theta$, the smoothness of $f$. Whatever the true rate of convergence
might be, Example~\ref{EX:newton-discrete} (model II) demonstrated
that this rate is unsatisfactory when the support of $f$ is
misspecified. For this reason, a modification of the algorithm that
better handles such cases would be desirable. 

Another question of interest goes back to the original motivation for
Newton's recursive algorithm. To an orthodox Bayesian, any method which
performs well should be at least \emph{approximately Bayes}. Stemming
from the fact that the recursive estimate and the nonparametric Bayes
estimate, with the appropriate Dirichlet process prior, agree when
$n=1$, Newton et al. \cite{nqz,newtonzhang,newton} claim that the
former should serve as a suitable approximation to the latter. Our
calculations in Section~\ref{SS:examples} disagree. In particular, we
see two examples in the finite case, one where the recursive estimate
is significantly better and the other where the Bayes estimate is
significantly better. A new question arises: if it is not an
approximation of the Dirichlet process prior Bayes estimate, \emph{for
what prior does the recursive estimate approximate the corresponding
Bayes estimate}?\looseness=1

Finally, it should be pointed out that our approach to the finite
mixture problem is somewhat less general than would be desirable. In
particular, we are assuming that the support of $f$ is within a \emph
{known} finite set of points. In general, however, what is known is
that the support of $f$ is contained in, say, a bounded interval. In
this case, a set of grid points $\Theta= \{\theta_1,\ldots,\theta
_m\}$ are chosen to approximate the \emph{unknown} support $\Theta^*
= \{\theta_1^*,\ldots,\theta_M^*\}$ of $f$. Newton's algorithm will
produce an estimate $f_n$ on $\Theta$ in this case, but it is
impossible to directly compare $f_n$ to $f$ since their supports
$\Theta$ and $\Theta^*$ may be entirely different. There is no
problem comparing the marginals, however. This leads us to the
following important conjecture, closely related to the so-called
$I$-projections in~\cite{csiszar,shyamalkumar}.
\begin{conj*}
Let $\Pi_{f_n}$ and $\Pi_f$ be the marginal densities corresponding
to $f_n$ on $\Theta$ and $f$ on $\Theta^*$, respectively. Then, as $n
\to\infty$,
\[
K(\Pi_f,\Pi_{f_n}) \to\inf_\varphi K(\Pi_f,\Pi_\varphi)
\quad \mbox{a.s.},
\]
where $\varphi$ ranges over all densities on $\Theta$.
\end{conj*}

Despite these unanswered practical and theoretical questions, the
strong performance of Newton's algorithm and the N$+$P algorithm in
certain cases and, more importantly, their computational
cost-effective\-ness, make them very attractive compared to the more
expensive nonparametric Bayes estimate or the nonparametric MLE, and
worthy of further investigation.

\begin{appendix}

\section*{Appendix: Proofs}\label{S:appendix}

\subsection{\texorpdfstring{Proof of Theorem
\protect\ref{T:martin-ghosh}}{Proof of Theorem 3.4}}
\label{SS:app-2}

To prove the theorem, we need only show that the algorithm \eqref
{E:newton2} satisfies the conditions of\vadjust{\goodbreak} Theorem~\ref{T:sa}. First note
that $f_n$ is, for each $n$, a convex combination of points in the
\emph{interior} of $\Delta^d$ so no projection as
in~\eqref{E:gen-sa} is necessary. Second, the random variables $\beta_n$ in
assumption SA2 are identically zero so SA3 is trivially satisfied.

Let $\{u_n\}$ be a convergent sequence in $\Delta^d$, where $u_n =
(u_n^1,\ldots,u_n^d)^\prime$. The limit
$u = (u^1,\ldots,u^d)^\prime=\break \lim_{n \to\infty} u_n$
also belongs to $\Delta$ so $h(u)$ is
well defined. To prove that $h = (h_1,\ldots,h_d)^\prime$ is
continuous, we show that $h_k(u_n) \to h_k(u)$ for each $k=1,\ldots,d$
as $n \to\infty$. Consider
\[
h_k(u_n) = \int\frac{p(x|\theta_k) u_n^k}{\Pi_{u_n}(x)} \Pi_f(x)
\, d\nu(x) - u_n^k.
\]
The integrand $p(\cdot|\theta_k) u_n^k/\Pi_{u_n}(\cdot)$ is
nonnegative and bounded $\nu$-a.e. for each $k$. Then by the bounded
convergence theorem we get
\[
\lim_{n \to\infty} h_k(u_n) = h_k(u), \quad k=1,\ldots,d.
\]
But $\{u_n\} \subset\Delta^d$ was arbitrary so $h$ is continuous.

Next, note that $H(x,f_n)$ is the difference of two points in $\Delta
^d$ and is thus bounded independent of $x$~and~$n$. Then SA1 holds trivially.

Finally, we show that $f$ is globally asymptotically stable for the ODE
$\dot{\varphi} = h(\varphi)$ in $\Delta^d$. Note that
$\sum_{i=1}^d \dot{\varphi}^i = \sum_{i=1}^d h_i(\varphi) = 0$
so the trajectories lie on
the connected and compact $\Delta^d$. Let $\ell(\varphi)$ be the KL
divergence, $\ell(\varphi) = \sum_{k=1}^d f^k \log(f^k/\varphi^k)$. We
claim\break that $\ell$ is a strong Lyapunov function for
$\dot{\varphi} = h(\varphi)$ at~$f$. Certainly $\ell(\varphi)$
is positive definite. To
check the differentiability condition, we must show that $\ell(\varphi)$
has a well-defined gradient around $f$, even when $f$ is on the
boundary of $\Delta^d$. Suppose, without loss of generality, that
$f^1,\ldots,f^s$ are positive, $1 \leq s \leq d$, and the remaining
$f^{s+1},\ldots,f^d$ are zero. By definition, $\ell(\varphi)$ is
constant in $\varphi^{s+1},\ldots,\varphi^d$ and, therefore, the partial
derivatives with respect to those $\varphi$'s are zero. Thus, for any
$1\leq s \leq d$ and for any $\varphi$ such that
$\ell(\varphi) <\infty$,
the gradient can be written as 
%
\begin{equation}\label{E:grad}
\nabla\ell(\varphi) = -(r^1,r^2,\ldots,r^d)' + r^sI_s',
\end{equation}
where $r^k = f^k/\varphi^k$ and $I_s$ is a vector whose first $s$
coordinates are one and last $d-s$ coordinates are zero. The key point
here is that the gradient of $\ell(\varphi)$, for~$\varphi$
restricted to
the boundary which contains $f$, is exactly 
\eqref{E:grad}. We can, therefore, extend the definition of
$\nabla\ell(\varphi)$ continuously to the boundary if need be.

Given that $\nabla\ell(\varphi)$ exists on all of $\Delta^d$, the time
derivative of $\ell$ along $\varphi$ is
%
\begin{eqnarray}\label{E:time-deriv}
\qquad
\dot{\ell}(\varphi) & =& \nabla\ell(\varphi)^\prime h(\varphi)
\nonumber \\
& = & \int\frac{\Pi_f(x)-\Pi_{\varphi}(x)}{\Pi_{\varphi}(x)}
\nabla\ell(\varphi)^\prime P_x \varphi\,d\nu(x)\hspace*{-5pt}
\\
& = & 1-\int\frac{\Pi_f}{\Pi_{\varphi}} \Pi_f \,d\nu.
\nonumber
\end{eqnarray}
It remains to show that $\dot{\ell}(\varphi) = 0$ iff $\varphi= f$.
Applying Jensen's inequality to $y \mapsto y^{-1}$ in
\eqref{E:time-deriv} gives
%
\begin{eqnarray}\label{E:jensens}
\dot{\ell}(\varphi) & =& 1-\int_{\mathcal{X}}
\biggl( \frac{\Pi_\varphi}{\Pi_f}  \biggr)^{-1} \Pi_f \,d\nu
\nonumber\\[-8pt]
\\[-8pt]
& \leq& 1-  \biggl( \int_{\mathcal{X}} \frac{\Pi_{\varphi}}{\Pi_f}
\Pi_f \, d\nu \biggr)^{-1} = 0,
\nonumber
\end{eqnarray}
where equality can hold in \eqref{E:jensens} iff $\Pi_{\varphi} =
\Pi_f$ $\nu$-a.e. We assume the mixtures are identifiable, so this
implies $\varphi= f$. Therefore, $\dot{\ell}(\varphi) = 0$ iff
$\varphi=f$, and we have shown that $\ell$ is a strong Lyapunov function on
$\Delta^d$. To prove that $f$ is a globally asymptotically stable
point for $\dot{\varphi}=h(\varphi)$, suppose that $\varphi(t)$ is a
solution, with $\varphi(0)=f_0$, that does not converge to $f$. Since
$\ell$ is a strong Lyapunov function, the sequence $\ell(\varphi(t))$,
as $t \to\infty$, is bounded, strictly decreasing and, thus, has a
limit $\lambda> 0$. Then the trajectory $\varphi(t)$ must fall in the set
\[
\Delta^* = \{\varphi\in\Delta^d\dvtx \lambda\leq\ell(\varphi)
\leq \ell (f_0)\}
\]
for all $t \geq0$. In the case $f \in\mbox{int}(\Delta^d)$,
$\ell(\varphi) \to\infty$ as $\varphi\to\partial\Delta$, so the set
$\Delta^*$ is compact (in the relative topology). If $f \in\partial\Delta^d$, then
$\Delta^*$ is not compact but, as shown above, $\dot{\ell}(\varphi)$
is well defined and continuous there. In either case, $\dot{\ell}$ is
continuous and bounded away from zero on $\Delta^*$, so
\[
\sup_{\varphi\in\Delta^*} \dot{\ell}(\varphi) = -L < 0.
\]
Then, for any $\tau\geq0$, we have
\[
\ell(\varphi(\tau)) = \ell(f_0) + \int_0^\tau\dot{\ell}
(\varphi(s))\, ds \leq\ell(f_0) - L\tau.
\]
If $\tau> \ell(f_0)/L$, then $\ell(\varphi(\tau)) < 0$, which is a
contradiction. Therefore, $\varphi(t) \to f$ for all initial conditions
$\varphi(0)=f_0$, so $f$ is globally asymptotically stable. Theorem
\ref{T:sa} then implies $f_n \to f$ a.s.

\subsection{\texorpdfstring{Proof of Theorem \protect\ref{T:martin-ghosh2}}{Proof of Theorem 4.1}}
\label{SS:NplusP-proof}

The proof of the theorem requires the following lemma, establishing a
Lipschitz-type bound on the error terms $\beta_n$ in \eqref{E:beta1}.
Its proof follows immediately from NP4 and the Mean Value Theorem.
\begin{lem}\label{L:lipschitz}
Under the assumptions of Theorem~\ref{T:martin-ghosh2}, there exists a
number $A \in(0,\infty)$ such that
\[
\|\beta_n\| \leq A \mathbb{E}(|\xi_n - \xi| \mid\mathscr{F}_{n-1}).
\]
\end{lem}
\begin{pf*}{Proof of Theorem \ref{T:martin-ghosh2}}
The map $h$ in \eqref{E:np-h} has $k$th component
\begin{eqnarray*}
h_k(\varphi) & =& \int H(x;\varphi,\xi) \Pi_{f,\xi}(s)\,d\nu^r(x)
\\
& = & \int\frac{\Pi_{f,\xi}(x)}{\Pi_{\varphi,\xi}(x)}
p(x|\theta_k,\xi) \varphi^k \,d\nu^r(x) - \varphi^k,
\end{eqnarray*}
where $\Pi_{f,\xi}(x) = \sum_k \overline{p}(x|\theta_k,\xi)f^k$
is the
marginal density of $x$ and $\nu^r$ is the product measure on
$\mathcal{X}^r$.
Notice that this $h$, which does not depend on the estimate $\xi_n$,
is \emph{exactly the same} as the $h$ in \eqref{E:h-function}.
Therefore, the continuity and stability properties derived in the proof
of Theorem \ref{T:martin-ghosh} are valid here as well. All that
remains is to show that the $\beta_n$'s in \eqref{E:beta1} satisfy
SA3 of Theorem~\ref{T:sa}.

By the SLLN, $\xi_n$ belongs to $\Xi_0$ for large enough $n$ so we
can assume, without loss of generality, that no projection is
necessary. Let $S_n = Z_1 + \cdots+ Z_n$, where the $Z_i =
v^{-1}(\hat{\xi}^{(i)}-\xi)$ and $v^2$ is the variance of $\hat{\xi}
^{(i)}$. Then $|\xi_n-\xi| = cn^{-1}|S_n|$, where $c > 0$ is a
constant independent of $n$. Since $S_n$ is a sum of i.i.d. random
variables with mean zero and unit variance, the Law of Iterated
Logarithm states that
%
\begin{equation}\label{E:lil1}
\limsup_{n \to\infty}  \{ |S_n|/u(n)  \} = 1 \quad\mbox{a.s.}
\end{equation}
Now, by Lemma \ref{L:lipschitz} and \eqref{E:lil1} we have
\[
\|\beta_n\| \leq Ac n^{-1} \mathbb{E}(|S_n| \mid\mathscr{F}_{n-1})
= O(n^{-1}u(n))
\]
and, therefore, $\sum_n w_n\|\beta_n\|$ converges a.s. by NP5.
Condition SA3 is satisfied, completing the proof.
\end{pf*}

\subsection{\texorpdfstring{Proof of Proposition
\protect\ref{P:normal-mix}}{Proof of Proposition 4.3}}
\label{SS:app-3}

To prove that the case of a location-mixture of normals with unknown
variance is covered by Theorem~\ref{T:martin-ghosh2}, we must show
that the function $H$, defined in~\eqref{E:H-function3}, satisfies NP4,
that is, that the partial derivatives
$\frac{\partial}{\partial\psi}H_k(s;\varphi,\psi)$ are bounded.
\begin{pf*}{Proof of Proposition \ref{P:normal-mix}}
Clearly each component $H_k$ of $H$, defined in \eqref{E:H-function3},
is differentiable with respect to $\psi\in\Sigma_0$ and, after
simplification,
%
\begin{eqnarray*}
\frac{\partial}{\partial\psi} H_k(s,\varphi,\psi)
&=& \frac{\varphi^k e^{-r\theta_k^2/2\psi} e^{rs\theta_k/\psi}}{2\psi^2}
\\
&&{}\cdot
\frac{\sum_j u_{kj}(s) \varphi^j e^{-r\theta_j^2/2\psi}
e^{rs\theta_j/\psi}}
{ [ \sum_j \varphi^j e^{-r\theta_j^2/2\psi}
e^{rs\theta_j/\psi}  ]^2},
\end{eqnarray*}
where (as $|s| \to\infty$)
%
\begin{equation}\label{E:linear-term}
\qquad
u_{kj}(s) = \theta_k^2 - \theta_j^2 + 2s(\theta_j-\theta_k) = O(|s|).
\end{equation}
This derivative is continuous on $s(\mathcal{X}^r) \times\Delta_0
\times
\Sigma_0$ and, since $\Delta_0$ and $\Sigma_0$ are compact, we know that
%
\begin{equation}
\label{E:lip1}
A_k(s) := \sup_{\varphi\in\Delta_0} \sup_{\psi\in\Sigma_0}
\biggl |\frac{\partial}{\partial\psi} H_k(s; \varphi,\psi)  \biggr|
\end{equation}
is finite for all $s \in s(\mathcal{X}^r)$ and for all $k$. By the
Mean Value Theorem,
\[
|H_k(s;\varphi,\psi)-H_k(s;\varphi,\sigma^2)|
\leq A_k(s) |\psi-\sigma^2|.
\]
It remains to show that $A_k(s)$ is bounded in $s$. For notational
simplicity, assume that $\varphi$ and $\psi$ are the values for which
the suprema in \eqref{E:lip1} are attained. Making a change of
variables $y = rs/\psi$ we can, with a slight abuse of notation, write
\[
A_k(y) \leq\frac{ C_k \varphi^k e^{y\theta_k}
\sum_j |u_{kj}(y)|\varphi ^j e^{y\theta_j}}
{ [ \sum_j \varphi^j e^{y\theta_j}  ]^2}.
\]
We must show that $A_k(y)$ is bounded as $|y| \to\infty$. Assume,
without loss of generality, that the $\theta$'s are arranged in
ascending order: $\theta_1 < \theta_2 < \cdots< \theta_d$.
Factoring out, respectively, $e^{y\theta_1}$ and $e^{y\theta_d}$, we
can write
\begin{eqnarray*}
A_k(y) & \leq& \frac{ C_k \varphi^k e^{y(\theta_k-\theta_1)} \sum_j
|u_{kj}(y)| \varphi^j e^{y(\theta_j-\theta_1)}}
{(\varphi^1)^2 +\sum_{j \neq1} \sum_{i \neq1} \varphi^j
\varphi^i e^{y(\theta_j-\theta_1)+y(\theta_i-\theta_1)}} ,
\\
A_k(y) & \leq& \frac{ C_k \varphi^k e^{y(\theta_k-\theta_d)}
\sum_j|u_{kj}(y)| \varphi^j e^{y(\theta_j-\theta_d)}}{(\varphi^d)^2
+ \sum_{j \neq d} \sum_{i \neq d} \varphi^j \varphi^i
e^{y(\theta_j-\theta _d)+y(\theta_i-\theta_d)}}.
\end{eqnarray*}
Note that since $\varphi\in\Delta_0$, each $\varphi^j$ is bounded away
from~0. If $y \to-\infty$, then the term
$e^{y(\theta_k-\theta_1)}\to0$ dominates the numerator of the first inequality, while the
denominator is bounded. Similarly, if $y \to+\infty$, then the term
$e^{y(\theta_k-\theta_d)} \to0$ dominates the numerator in the
second inequality, while the denominator is bounded. For the case $k=1$
or $k=d$, note that $|u_{11}(y)| = |u_{dd}(y)| = 0$, so the two
inequalities can still be applied and a similar argument shows $A_1$
and $A_d$ are also bounded. Therefore, $A_k(y)$ is bounded for each $k$
and the claim follows by taking $A$ to be $\max\{\sup_y A_k(y)\dvtx
1\leq k \leq d \} $.
\end{pf*}

\end{appendix}

\section*{Acknowledgments}

The authors thank Professors Chuanhai Liu and Surya T. Tokdar for
numerous fruitful discussions, as well as Professors Jim Berger and
Mike West, the Associate Editor and the two referees for their helpful comments.

\vspace*{-1.2pt}
\end{document}